\documentclass[pre,
onecolumn,
preprint,
superscriptaddress,
showpacs,
amsmath,
amssymb,
floatfix]{revtex4-1}

\usepackage{units}
\usepackage{psfrag}

\usepackage[dvips]{graphicx}

\newcommand{\e}{\varepsilon}

\newcommand{\w}{\omega}

\begin{document}

\title{Inter-community resonances in multifrequency ensembles of coupled oscillators}

\author{Maxim Komarov} 
\affiliation{Department of Physics and Astronomy, University of Potsdam,
  Karl-Liebknecht-Str 24, D-14476, Potsdam, Germany}
\affiliation{Department of Control Theory, Nizhni Novgorod State University,
  Gagarin Av. 23, 606950, Nizhni Novgorod, Russia}
\author{Arkady Pikovsky} 
\affiliation{Department of Physics and Astronomy, University of Potsdam, 
  Karl-Liebknecht-Str 24, D-14476, Potsdam, Germany}

\date{\today}

\begin{abstract}
We generalize the Kuramoto model of globally coupled oscillators to multifrequency 
communities. A situation when mean frequencies of two subpopulations are close to 
resonance 2:1 is considered in detail. We derive uniformly rotating solutions describing 
synchronization inside communities and between them. Remarkably, cross-coupling between the
frequency scales can promote synchrony even when ensembles are separately asynchronous. We 
also show that the transition to synchrony due to the cross-coupling is accompanied by a 
huge multiplicity of distinct synchronous solutions what is directly related to a 
multi-branch entrainment. On the other hand, for synchronous populations,
the cross-frequency coupling can destroy a phase-locking and lead to chaos of mean fields.  
\end{abstract}


\maketitle
\section{Introduction}
Models in the form of coupled oscillators are ubiquitous  in various scientific fields,
ranging from physics and chemistry~\citep{Wiesenfeld-Swift-95,*Wiesenfeld-Colet-Strogatz-96,%
*Wiesenfeld-Colet-Strogatz-98,*Kiss-Zhai-Hudson-02a,*Grollier-Cros-Fert-06,*georges:232504} 
to biology~\citep{Golomb-Hansel-Mato-01,%
*Breakspear-Heitmann-Daffertshofer-10, *Gonze-05, *Bordyugov-13}, 
as well as in some interdisciplinary applications~\citep{Eckhardt_et_al-07,*Neda_etal-00}. 
In many cases dynamics of oscillatory ensemble can be successfully studied in the
phase approximation ~\citep{Kuramoto-84,Pikovsky-Rosenblum-Kurths-01}.
When the coupling between the oscillators is relatively weak, one can neglect changes in 
the 
amplitude dynamics of natural limit cycles of the oscillators,
and describe the system in terms 
of the phases only. This technique is known as phase reduction, and
it represents, basically,  
one of the few rigorous mathematical approaches to study complex non-equilibrium nonlinear oscillatory 
dynamics.

The simplest setup here represents a globally coupled ensemble with weak interaction and 
relatively close natural frequencies. 
The phase reduction here leads to the system of globally coupled phase equations where 
interaction between the oscillators is described by $2\pi$ - periodic function of phase 
differences ~\citep{Kuramoto-84,Kuramoto-75,Izhikevich-07,*Izhikevich-00,%
Daido-93,*Daido-93a,*Daido-96,*Daido-95}.
The classical and well-studied Kuramoto-Sakaguchi model appears when one consider only the 
first Fourier mode in the interaction function what leads to simple sinusoidal coupling.
There is almost 40 years of intensive studies dedicated to explanation of bifurcations and 
dynamics in this model~\citep{Acebron-etal-05}.
A surprising recent result discovered a possibility of low-dimensional description of the 
classical Kuramoto model in terms of macroscopic order 
parameters~\citep{Watanabe-Strogatz-94,Ott-Antonsen-08,*Ott-Antonsen-09,%
Marvel-Mirollo-Strogatz-09,*Pikovsky-Rosenblum-08,*Pikovsky-Rosenblum-11}.
However, this reduction to low-dimensional systems does not imply 
simplicity of dynamical behavior.
In opposite, the authors \citep{Omelchenko-Wolfrum-12,*Omelchenko-Wolfrum-13} report on quite complicated phase transitions and bifurcations in the Kuramoto-Sakaguchi models.

The cases of  multi-harmonic coupling 
functions~\citep{Daido-95,Daido-96a,Daido-96,*Crawford-95,*Crawford-Davies-99,%
*Chiba-Nishikawa-11,*Hansel-93,*Ashwin_etal-07} appear to be more complicated and usually 
responsible for new dynamical effects in comparison to classical setup with purely 
sinusoidal function. 
In large ensembles the multi-harmonic case leads to appearance of so-called multi-branch 
entrainment modes with a huge multiplicity of possible synchronous 
solutions~\citep{Daido-95,Daido-96a,Komarov-Pikovsky-13a,*Komarov-Pikovsky-14}. The latter 
also leads to non-trivial noise-induced effects~\citep{Vlasov-Komarov-Pikovsky-15}. 

One of the directions in this growing 
theoretical field is dedicated to \textit{multi-frequency} 
oscillator communities. 
As it was mentioned before, the Kuramoto-type models were obtained under assumptions of 
weak coupling limit and closeness of natural oscillator frequencies.
However, when the distribution of the frequencies is huge in comparison to the
interaction 
strength, the phase reduction leads to another types of 
phase 
models ~\citep{Lueck-Pikovsky-11,Komarov-Pikovsky-13,Komarov-Pikovsky-11}.
A natural setup here implies existence of a certain number of oscillator 
subpopulations (communities), such that the frequencies inside each population are close, 
but differ 
significantly across the distinct communities. 
This situation is inspired by theoretical and experimental results 
from neuroscience~\citep{Buzsaki-06,*Rosjat-Popovych-Daun-14}, indicating that distinct 
interacting brain areas exhibit different natural oscillatory rhythms.

In this paper we consider a particular problem when distinct oscillatory communities have 
natural frequencies close to a high-order resonance.
First, we derive general phase equations for globally interacting ensembles and distinguish 
different types of resonant coupling which may appear in the system.
Next, we concentrate on the simplest case of two interacting population whose mean 
frequencies are close to an 2:1 resonance.
The aim of the paper is to demonstrate on this simplest example, what one can expect from 
the effects high-order resonances.
To describe the dynamics, we adopt the self-consistent approach developed in~\citep{Komarov-Pikovsky-14} for calculation of stationary order parameters for multi-harmonic coupling 
functions.
Our analysis will show that the model exhibits reach 
dynamical behavior including multi-branch entrainment (multiplicity) and chaotic collective oscillations. 
 
\section{Phase equations for resonantly coupled populations}
In this section we will present a general scheme of coupling in resonant, multifrequency
populations of oscillators. We will assume that each oscillator is described solely by its phase $\phi$, which satisfies the following equation
\[
\dot\phi=\omega+S(\phi)F
\]
where $\omega$ is oscillator's natural frequency, $S(\phi)$ is its phase response curve, and
$F$ is the force acting from other oscillators. In order to simplify notations, 
we from the beginning will consider a thermodynamic limit, where the number of 
units in all populations and subpopulations tends to infinity (although at the 
end we will also write the governing equations for a finite size case). We assume 
that the ensemble is divided into $M$ distinct subpopulations (we will use index $n$ 
for referring to them), around $M$ distinct 
mean frequencies $\overline{\omega}_n$. Additionally, there can be a small deviation 
from the mean frequency $\Delta$ (typically described by a unimodal distribution 
around zero). We now introduce slow phases, by writing explicitly
fast rotating terms $\sim  \overline{\omega}_n t$. In fact, we can also chose frequencies
of fast rotations
$\Omega_n$ to be close, but not exactly equal, to  $\overline{\omega}_n$. We will use this freedom to be able to make 
perfect averaging below. Our slow phases 
$\varphi_n(\Delta)=\phi_n-\Omega_n t$ satisfy equations
\begin{equation}
\dot\varphi_n(\Delta)=\Delta+S_n(\Omega_nt+\varphi_n)F_n
\label{eq:slph}
\end{equation}
where now individual mismatches $\Delta$ for the group $n$ are distributed 
generally asymmetrically, with some small shift $\sim \overline{\omega}_n-\Omega_n$. 

Next, we assume that coupling between the groups and inside each group is due to
mean fields only. These mean fields for each subpopulation are represented by generalized order parameters
\[
Z_k^{(n)}=\langle e^{ik(\Omega_n t+\varphi_n)}\rangle=\overline{Z}_k^{(n)}e^{ik\Omega_n t}
\]
where averaging is over the distribution of the slow phases following from  
\eqref{eq:slph} and over the distribution of $\Delta$. The introduced order parameters 
$\overline{Z}$ are slow functions of time as they are defined via the slow phases:
\begin{equation}
\overline{Z}k^{(n)}=\langle e^{ik\varphi_n}\rangle
\label{eq:slop}
\end{equation}
In general, the force acting on the oscillators of the group $n$ is from all other groups, and is a nonlinear function of order parameters, which one can expand in powers of them.
We, however, in this paper will restrict ourselves to the linear coupling only, i.e. we will assume that $F_n$ is a linear function of order parameters:
\begin{equation}
F_n(Z_k^{(1)},Z_k^{(2)},\ldots)=\sum_{k,m}h_{n,k}^{(m)} Z_k^{(m)}
=\sum_{k,m}h_{n,k}^{(m)} \overline{Z}_k^{(m)}e^{ik\Omega_m t}
\end{equation}

Representing the phase response function $S_n$ as a Fourier series
\[
S_n(\phi)=\sum_p s_{np} e^{ip\phi}
\]
and substituting this in Eq.~\eqref{eq:slph}, we obtain
\begin{equation}
\begin{aligned}
\dot\varphi_n(\Delta)&=\Delta+\sum_p s_{np} e^{ip\varphi}e^{ip\Omega_n t}
\left[\sum_{k,m}h_{n,k}^{(m)} \overline{Z}_k^{(m)}e^{ik\Omega_m t} 
\right]=\\
&=\Delta+\sum_{p,k,m} s_{np} h_{n,k}^{(m)}  \overline{Z}_k^{(m)} e^{ip\varphi}e^{i(p\Omega_n +ik\Omega_m )t} 
\end{aligned}
\label{eq:slph2}
\end{equation}

Now one has to perform averaging of Eq.~\eqref{eq:slph2}, to reveal evolution of the slow 
phase. The fast terms on the r.h.s. are those containing explicit time dependence with one 
of the frequencies $\Omega_n$ or with a combination of them. Such a combination can be small,
this is exactly the case of a resonance that is of special interest for us.  Here, we use the
freedom in the choice of particular values of $\Omega_n$, to make the resonance exact. This 
means that some combination of frequencies $\Omega_n$ vanishes exactly. Performing 
averaging means just keeping these terms 
on the r.h.s. of Eq.~\eqref{eq:slph2}, and neglecting all 
other containing explicit time dependence.

Expansion~\eqref{eq:slph2} can be treated in many setups of particular 
resonant conditions, we describe here some evident cases:
\begin{itemize}
\item One population of oscillators. In this case only one  frequency $\Omega$ exists. 
Here the only terms surviving the averaging are those with $p+k=0$, 
this leads to the Daido model~\cite{Daido-93,*Daido-93a,*Daido-96,*Daido-95}. 
\item Two subpopulations. Here the main interest is in the resonance of two frequencies 
$\Omega_1,\Omega_2$. The simplest case is just the second-harmonic resonance: $\Omega_2=2\Omega_1$.
In this case only those cross-population coupling terms with $p+2k=0$ survive. 
Similarly, for high-order 
resonances like $a\Omega_2=b\Omega_1$ (with integer $a,b$) the terms with $ap+bk=0$ 
contribute.
\item More than two subpopulations. One can see from~\eqref{eq:slph2}, that in the case of 
linear coupling, there is no direct interaction involving more that two subpopulations. So 
the resulting coupling is a combination of terms stemming from pairwise resonances. 
\end{itemize}

We restrict ourself in this paper to the simplest case of two resonant subpopulations with $\Omega_2=2\Omega_1$.
As described above, after averaging only terms where combinations $\sim (\Omega_{1,2}-\Omega_{1,2})$ and $\sim(\Omega_2-2\Omega_1)$ appear, survive, for the interaction within one and between subpopulations, respectively:
\begin{equation}
\begin{aligned}
\dot\varphi_1(\Delta)
&=\Delta+\sum_{k} s_{1,-k} h_{1,k}^{(1)}  \overline{Z}_{k}^{(1)} e^{-ik\varphi_1} +\sum_{k} s_{1, -2k} h_{1,k}^{(2)}  \overline{Z}_k^{(2)} e^{-i2k\varphi_1} \\
\dot\varphi_2(\Delta)
&=\Delta+\sum_{k} s_{2,-k} h_{2,k}^{(2)}  \overline{Z}_{k}^{(2)} e^{-ik\varphi_2}+\sum_{k} s_{2,-k} h_{2,2k}^{(1)}  \overline{Z}_{2k}^{(1)} e^{-ik\varphi_2} 
\end{aligned}
\label{eq:slph21}
\end{equation}
We now insert here the definition of the slow order parameters~\eqref{eq:slop} and obtain
\begin{equation}
\begin{aligned}
\dot\varphi_1(\Delta)
&=\Delta+\langle\sum_{k} s_{1,-k} h_{1,k}^{(1)}   e^{ik(\tilde\varphi_1-\varphi_1)}\rangle 
+\langle \sum_{k} s_{1, -2k} h_{1,k}^{(2)}   e^{ik(\tilde\varphi_2-2\varphi_1)}\rangle \\
\dot\varphi_2(\Delta)
&=\Delta+\langle \sum_{k} s_{2,-k} h_{2,k}^{(2)}  e^{ik(\tilde\varphi_2-\varphi_2)}\rangle+\langle \sum_{k} s_{2,-k} h_{2,2k}^{(1)}  
e^{ik(2\tilde\varphi_1-\varphi_2)}\rangle 
\end{aligned}
\label{eq:slph3}
\end{equation}
where averaging is over variables with tilde. Now we can define effective
coupling functions inside
subpopulations $f_{11},f_{22}$ and coupling functions across subpopulation $f_{12},f_{21}$ as
\begin{equation}
\begin{aligned}
&f_{11}(\phi)=\sum_{k} s_{1,-k} h_{1,k}^{(1)}e^{ik\phi}\qquad
f_{22}(\phi)= \sum_{k} s_{2,-k} h_{2,k}^{(2)}  e^{ik\phi}\\
&f_{12}(\phi)=\sum_{k} s_{1, -2k} h_{1,k}^{(2)}   e^{ik\phi}\qquad
f_{21}(\phi)=\sum_{k} s_{2,-k} h_{2,2k}^{(1)}  e^{ik\phi}
\end{aligned}
\end{equation}
Now we can formulate equations for finite populations, replacing $\langle\rangle$ by 
corresponding sums. We assume that subpopulations 1 and 2 have $N_1$ and $N_2$ units, 
respectively. Furthermore, one can now also transform back to the original fast phases, 
because in the averaged formulation the absolute 
values of the frequencies do not play any r\^ole. Denoting the phases in the subpopulation at a 
smaller frequency (we will also call it the first subpopulation below)
as $\phi_p$, and the phases in the  subpopulation at a larger frequency 
(referred hereafter as the second subpopulation) as 
$\psi_p$, we get
\begin{equation}
\begin{aligned}
\dot\phi_q&=\omega_q+\frac{1}{N_1}\sum_{k=1}^{N_1} f_{11}(\phi_k-\phi_q)+\frac{1}{N_2}
\sum_{p=1}^{N_2} f_{12}(\psi_p-2\phi_q)\\
\dot\psi_q&=\nu_q+\frac{1}{N_2}\sum_{k=1}^{N_2} f_{22}(\psi_k-\psi_q)+
\frac{1}{N_1}\sum_{p=1}^{N_1} f_{21}(2\phi_p-\psi_q)
\end{aligned}
\end{equation}
where we also have split notations for frequencies in two subpopulations. 
This system is a generalization of the Daido 
model~\cite{Daido-93,*Daido-93a,*Daido-96,*Daido-95} to two resonantly coupled ensembles. 
Below we
will consider the case where coupling functions $f$ contain the first harmonics only; this 
will correspond to the Kuramoto-Sakaguchi-type coupling. In this case each coupling 
function is determined by two parameters, the amplitude and the phase shift. One of the 
phase shifts in the cross-coupling can be set to zero by shifting all the phases in one 
subpopulation with respect to another one. Thus, our coupling functions will be:
\[
f_{11}(x)=\e_1\sin(x-\alpha_1),\quad
f_{22}(x)=\e_2\sin(x-\alpha_2),\quad
f_{12}(x)=\gamma_1\sin(x-\beta),\quad
f_{22}(x)=\gamma_2\sin x
\]

Next, we fix the distributions of the frequencies. As after the averaging the system is 
invariant under transformation $\phi\to \phi+2At$, $\psi\to \psi+At$ for arbitrary $A$, we 
can set the average value of the natural frequencies in the 
first subpopulation $\phi$ to zero, the average frequency $\delta$ in the second 
subpopulation is the relevant parameter responsible for the mismatch. We will assume the
frequencies to be distributed according to Lorentzian distributions, with equal
widths. Because we still have a freedom of changing the time scale, 
we will assume that this width is one:
\begin{equation}
g_1(\omega)=\frac{1}{\pi(\omega^2+1)}\qquad g_2(\nu)=
\frac{1}{\pi((\nu-\delta)^2+1}
\label{eq:distr}
\end{equation}
The resulting microscopic system of oscillators to be considered below reads
\begin{equation}
\begin{aligned}
\dot{\phi}_n =  \omega_n+\frac{\varepsilon_1}{N_1} \sum_{k=1}^{N_1} 
\sin(\phi_k-\phi_n-\alpha_1)+\frac{\gamma_1}{N_2} \sum_{k=1}^{N_2} 
\sin(\psi_k-2\phi_n-\beta)\\
\dot{\psi}_m = \nu_m + \frac{\varepsilon_2}{N_2}\sum_{k=1}^{N_2} 
\sin(\psi_k-\psi_m-\alpha_2) + \frac{\gamma_2}{N_1} 
\sum_{k=1}^{N_1}\sin(2\phi_k-\psi_m)
\end{aligned}
\label{eq:bkm_micro}
\end{equation}
with frequencies defined according to the distributions~\eqref{eq:distr}. 

We now also write down the basic equations in the thermodynamic limit. Here three complex order parameters $\mathbf{X}_1,\mathbf{X}_2,\mathbf{Y}$ appear defined as
\begin{equation}
\begin{aligned}
\mathbf{X}_{k}&=X_ke^{i\Theta_k}=\langle e^{ik\phi}\rangle= \iint d\phi d\omega\;g_1(\w)\rho(\phi|\omega) e^{ik\phi},\qquad
k=1,2\;,\\
\mathbf{Y}&=Ye^{i\Theta_y}=\langle e^{i\psi}\rangle= \iint d\psi d\nu\;g_2(\nu)\rho(\psi|\nu) e^{i\psi}
\end{aligned}
\label{eq:opXY}
\end{equation}
while equations for the phases are
\begin{equation}
\begin{aligned}
\dot\phi&=\omega+\e_1X_1\sin(\Theta_1-\phi-\alpha_1)+\gamma_1Y\sin(\Theta_y-2\phi_1-\beta)\\
\dot\psi&=\nu+\e_2Y\sin(\Theta_y-\psi-\alpha_2)+\gamma_2X_2\sin(\Theta_2-\psi)
\end{aligned}
\label{eq:pheqthl}
\end{equation}

The formulated system of equation will be subject of our analysis below, where we will 
concentrate on main dynamical effects caused by resonant cross-coupling. In numerical simulations we will use microscopic equations~\eqref{eq:bkm_micro}, while in the theoretical 
construction the thermodynamic limit formulation~(\ref{eq:opXY},\ref{eq:pheqthl}) will be used.

 \section{Self-consistent solutions in the thermodynamic limit}

 Here we will present the self-consistent scheme allowing us to find stationary (or, more generally, uniformly rotating) synchronous solutions of the system (\ref{eq:opXY},\ref{eq:pheqthl}). 
 
 \subsection{Ott-Antonsen ansatz for the second subpopulation}
 The problem partially simplifies by the observation, that for the subpopulation $\psi$ at the double frequency the Ott-Antonsen ansatz~\cite{Ott-Antonsen-08,*Ott-Antonsen-09} can be applied. Indeed,
 the second of eqs.~\eqref{eq:pheqthl} can be rewritten as
 \begin{equation}
 \dot\psi=\nu+\text{Im}(\mathbf{H}(t)e^{-i\psi}),\quad 
 \mathbf{H}=\e_2e^{-i\alpha_2}\mathbf{Y}+\gamma_2\mathbf{X}_2
 \label{eq:psi1}
 \end{equation}
 According to the Ott-Antonsen theory, equation for the order parameter $\mathbf{Y}$ obeys
 (under some additional assumptions which we assume to be satisfied here),
 in the case of a Lorentzian distribution~\eqref{eq:distr}, an ODE
 \begin{equation}
\dot{\mathbf{Y}} = \mathbf{Y}(i\delta-1) -\frac{1}{2}(\mathbf{H}^*\mathbf{Y}^2-\mathbf{H})=
\mathbf{Y}(i\delta-1)-\frac{\varepsilon_2\mathbf{Y}}{2}(e^{-i\alpha_2}|\mathbf{Y}|^2-e^{i\alpha_2}) - \frac{\gamma_2}{2}(\mathbf{X}_2^{*}\mathbf{Y}^2-\mathbf{X}_2)
\label{eq:oa}
\end{equation}

\subsection{Uniformly rotating ansatz}

We now construct solution for the ensemble of oscillators $\varphi$. Here we cannot use
the Ott-Antonsen ansatz, because the latter is only applicable for the driving terms possessing one harmonics of the phase, like in~\eqref{eq:psi1}. The equations for $\varphi$ possesses both the first and the second harmonics. 
In order to find stationary values of the mean fields, we will adapt the self-consistent scheme developed in Refs.~\citep{Komarov-Pikovsky-13a,*Komarov-Pikovsky-14} for the deterministic bi-harmonic Kuramoto model (for the noisy case a similar method can be used, see~\citep{Vlasov-Komarov-Pikovsky-15,Komarov-Gupta-Pikovsky-14}).

In this self-consistent approach one finds uniformly rotating distributions, i.e. distributions that are stationary in a rotating reference frame. Let us denote the
frequency of this frame $\Omega$, it will be determined self-consistently as a result of the calculations. According to this, we introduce constant phases of the order parameters
\begin{equation}
\Theta_1=\Omega t,\quad \theta_2=\Theta_2-2\Omega t,\quad \theta_y=\Theta_y-2\Omega t
\label{eq:var_tr}
\end{equation}
(here the phase shift of the first order parameter $\mathbf{X}_1$ is set to zero, this can
be always done by the time shift). 
Also, we introduce a new phase variable
\[
\varphi=\phi-\Omega t+\alpha_1
\]
distribution of which is expected to be stationary. This variable obeys
\begin{equation}
\dot{\varphi} = \omega-\Omega+\e_1 X_1\sin(-\varphi)+\gamma_1 Y\sin(\theta_y + 2\alpha_1-\beta-2\varphi)
\label{eq:bkm}
\end{equation}

\subsection{Stationary solution in a parametric form}

To proceed with self-consistent solution, it is convenient to introduce four auxiliary
parameters $\{R,\ u,\ v,\ z\}=\mathbf{P}$ in the following way:
\begin{equation}
\e_1 X_1=R\sin{u},\ \gamma_1 Y = R\cos{u},\ \Omega = zR,\ v = \theta_y+2\alpha_1-\beta
\label{eq:params}
\end{equation}

Now (\ref{eq:bkm}) takes the following form:
\begin{equation}
\dot{\varphi} =R\left(x - z -\sin{u}\sin{\varphi}-\cos{u} \sin(2\varphi-v)\right) = R\left(x - z - h(u,v,\varphi)\right)
\label{eq:bkm_first}
\end{equation}
We denoted $x = \omega/R$ and $h(u,v,\varphi)=\sin u \sin \varphi + \cos
u\sin(2\varphi-v)$.
At some constant values of parameters $\mathbf{P}$ in (\ref{eq:bkm_first}), at each
value of $x$ one can find stationary distribution function $\rho(\varphi|x,\mathbf{P})$, and then calculate the corresponding complex order parameters:
\begin{equation}
\begin{aligned}
X_1 &= e^{-i\alpha_1}R\iint\rho(\varphi|x,\mathbf{P}) 
e^{i\varphi}g(Rx)dx d\varphi = e^{i(\theta_1-\alpha_1)}R F_1(\mathbf{P}) e^{iQ_1(\mathbf{P})}\\
X_2e^{i\theta_2} &= e^{-i2\alpha_1}R
\iint\rho(\varphi|x,\mathbf{P}) e^{i2\varphi}g(Rx)dx d\varphi = 
e^{-i2\alpha_1}R F_2(\mathbf{P}) e^{iQ_2(\mathbf{P})}\\
F_{m}(\mathbf{P}) e^{iQ_{m}(\mathbf{P})} & \equiv \iint 
dxd\varphi\rho(\varphi|x,\mathbf{P})
e^{im\varphi}g(Rx)\;,\qquad
m=1,2\;.
\end{aligned}
\label{eq:int}
\end{equation}

\begin{figure}[!htb]
\centerline{(a)\includegraphics[width=0.49\columnwidth]{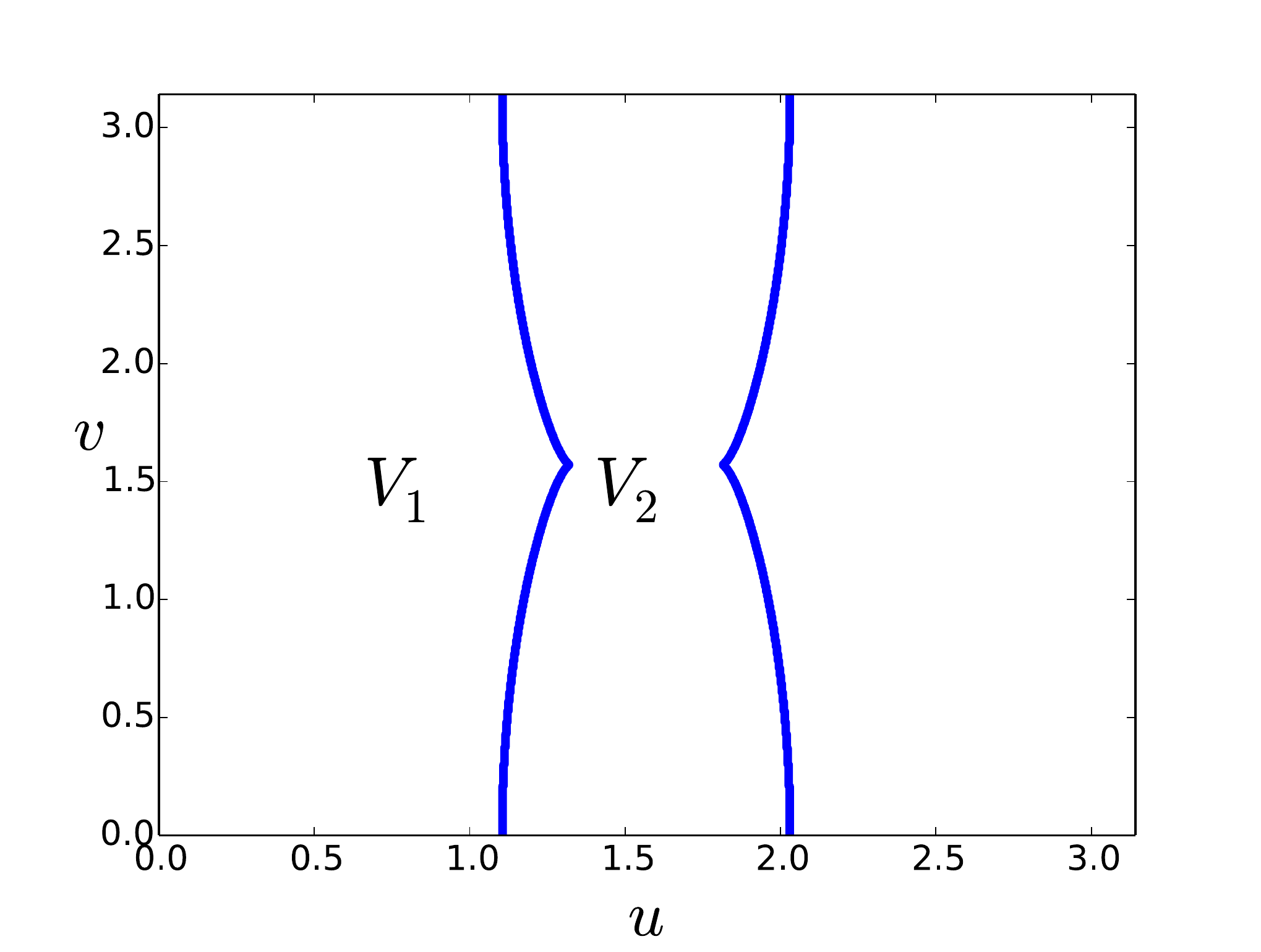} }
\centerline{(b)\includegraphics[width=0.49\columnwidth]{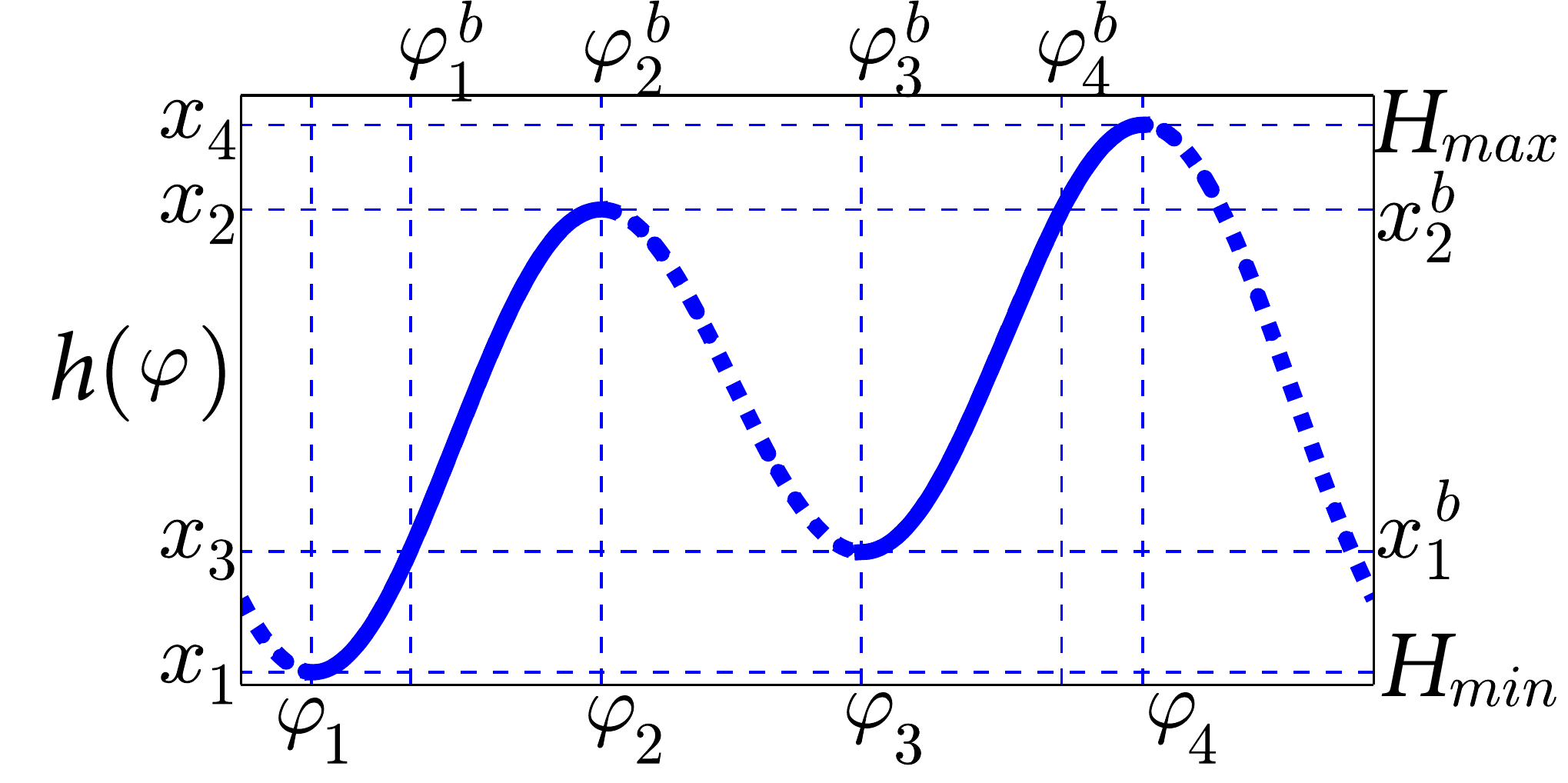} (c)\includegraphics[width=0.49\columnwidth]{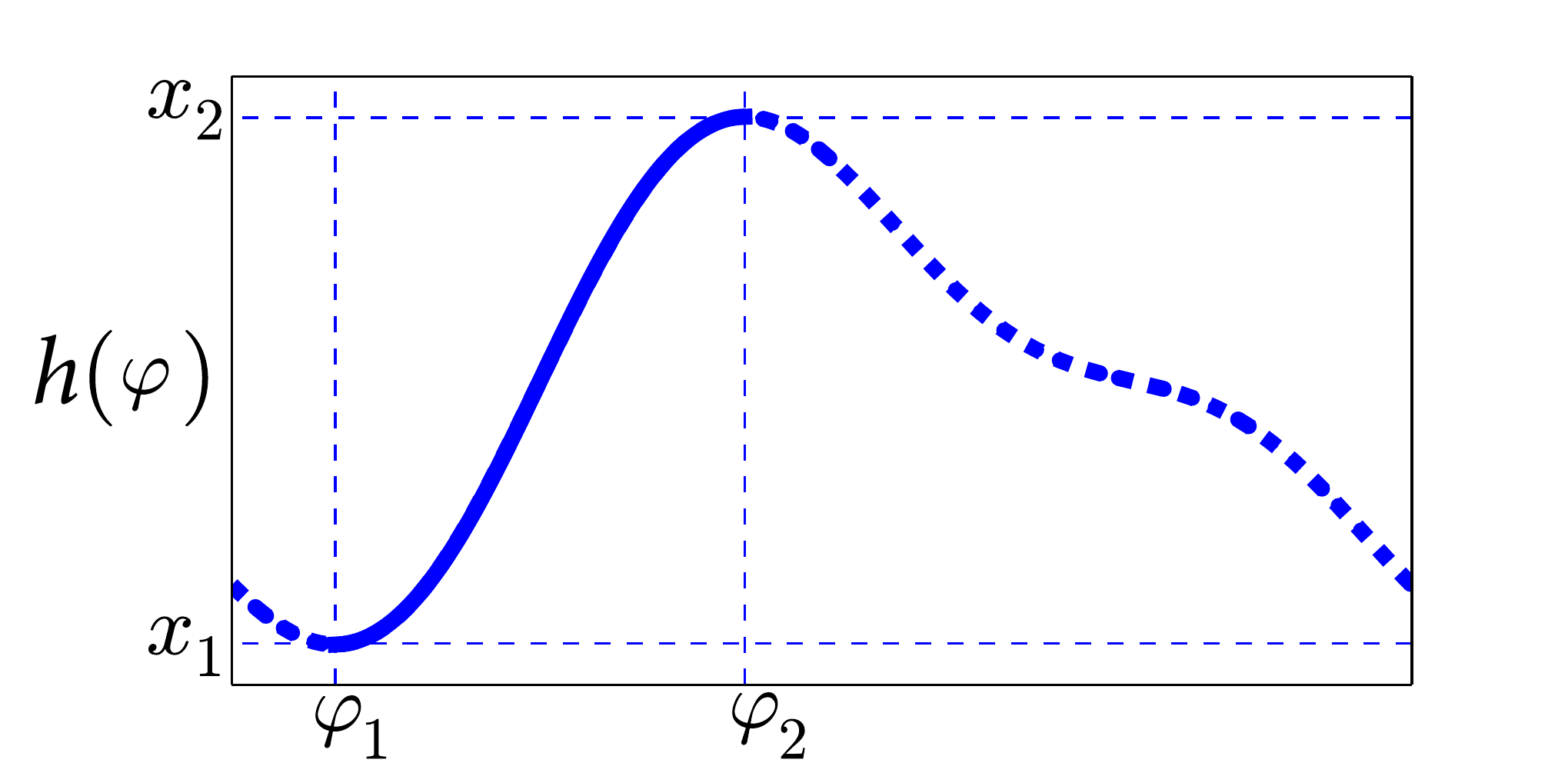}}
\caption{(a) Regions $V_1$ and $V_2$ in the plane of
parameters $(u,v)$: 
Domain $V_1$ corresponds to a double-well form of function $h(u,v,\varphi)$
(Fig.~\ref{fig:locked_ph}(b,d)), while in $V_2$ 
$h(u,v,\varphi)$ has a single-well form
like sown in Fig.~\ref{fig:locked_ph}(c). 
(b) Example of function $h(u,v,\varphi)$ with 4 extrema is presented. There are two
stable branches (solid curves) for stationary phases of locked oscillators. The
 left branch $\varphi=\Psi_1(x,\mathbf{P})$ is larger than the right one
$\varphi=\Psi_2(x,\mathbf{P})$. $(\varphi_{1,2},x_{1,2})$ denote coordinates of the
extrema corresponding to the  branch $\Psi_1$, while
$(\varphi_{3,4},x_{3,4})$ denotes extrema at $\Psi_2$. 
 In the domain $h(\varphi)\in[x^b_1,x^b_2]$ there is a bistability on the microscopic level: 
in this domain the oscillators can be locked either on the branch $\Psi_1$ in the 
range $\varphi \in [\varphi^b_{1},\varphi^b_2]$ or on the branch $\Psi_2$ in the range $\varphi \in [\varphi^b_{3},\varphi^b_4]$.
(c) Example of function $h(u,v,\varphi)$ with only two extrema and one stable
branch $\varphi=\Psi_1(x,\mathbf{P})$  (solid curve). 
}
\label{fig:locked_ph}
\end{figure}

Our next goal is to calculate the integrals $F_m(\mathbf{P})$, for this we need 
to find, using the dynamical equation (\ref{eq:bkm_first}), 
the stationary distribution function $\rho(\varphi|x,\mathbf{P})$.
Let $H_{min}$ and $H_{max}$ denote the global minimum and the 
global maximum of function
$h(u,v,\varphi)$, correspondingly (Fig.\ref{fig:locked_ph}(b)). 
All the oscillators can be separated into locked ones (for
$H_{max}\geq|x-z|\geq H_{min}$) or rotating, unlocked 
ones ($x-z>H_{max}$ or $x-z<H_{min}$). 
The distribution function of rotating oscillators (index $r$) is inversely 
proportional to
their phase velocity:
\begin{equation}
\rho_r(\varphi|x,\mathbf{P}) =  
\frac{C(x)}{|x-z-h(\varphi,u,v)|}\;,
\label{eq:rdist}
\end{equation}
where $C(x)$ is the normalization constant:
\[
C(x) = \frac{1}{\int_0^{2\pi}\frac{d\varphi}{|x-z-y|}}.
\]

The stationary phases of locked oscillators (index $l$)
can be found from the following
relation:
\begin{equation}
x - z = h(u,v,\varphi)\;.
\label{eq:locked_ph}
\end{equation}

When finding $\varphi$ as a function of $x$ for non-rotating 
(locked, index $l$) phases, we have 
to satisfy an additional stability
condition $\frac{\partial h(u,v,\varphi)}{\partial \varphi}>0$ that follows from
the dynamical equation
(\ref{eq:bkm_first}).
In the $(u,v)$ plane there are two regions $V_1$ and $V_2$
(Fig.~\ref{fig:locked_ph}(a)) with qualitatively different properties
of the system (\ref{eq:bkm_first}) and different types of distribution function
$\rho_l(\varphi|x,\mathbf{P})$, correspondingly:
\paragraph{$\{u,v\}\in V_1$} In this case function $h(u,v,\varphi)$ has a
double-well form like shown in Fig.\ref{fig:locked_ph}(b). 
According to (\ref{eq:bkm_first}), oscillators can be located on two possible
stable branches highlighted by solid curves in Fig.\ref{fig:locked_ph}(b): the
first branch is $\varphi = \Psi_1(x,\mathbf{P})$ in the range $\varphi\in
[\varphi_1,\varphi_2]$ and another branch is $\varphi = \Psi_2(x,\mathbf{P})$ for
$\varphi\in [\varphi_3,\varphi_4]$.
Here and below we assume $\Psi_1(x,\mathbf{P})$ to be the biggest stable branch.
In the range $(x-z)\in (x^b_1,x^b_2)$ (Fig.~\ref{fig:locked_ph}(b)) there is an
area of bistability on the microscopic level: the oscillators with the 
same natural
frequency $x$ can be locked at two 
different phases $\Psi_1(x,\mathbf{P})$ and
$\Psi_2(x,\mathbf{P})$.
Therefore, the distribution function has the following form:
\begin{equation}
\rho_{l}(\varphi|x,\mathbf{P})=\left\{
\begin{array}{l}
(1-S(x))\delta(\psi-\Psi_1(x,\mathbf{P}))+S(x)\delta(\varphi-\Psi_2(x,\mathbf{P}
))\\ 
\text{for $(x-z)\in[x^b_1,x^b_2]$}\;,\\
\delta(\varphi-\Psi_1(x,\mathbf{P}))\ \ \ \text{for $(x-z)\in [x_1,x_2]\setminus
[x_1^b,x_2^b]$}\;,\\
\delta(\varphi-\Psi_2(x,\mathbf{P}))\ \ \ \text{for $(x-z)\in [x_3,x_4]\setminus
[x_1^b,x_2^b]$}\;.\\
\end{array}
\right.
\label{eq:distr_func_1}
\end{equation}
Here $0\leq S(x)\leq 1$ is an indicator function describing the 
redistribution over the
stable brunches; this function is arbitrary.

\paragraph{$\{u,v\}\in V_2$} In the second case, 
 function $h(u,v,\psi)$ has
only two extrema (Fig.~\ref{fig:locked_ph}(c)) and there is only one stable
branch $\varphi = \Psi_1(x,\mathbf{P})$. The distribution function is:
\begin{equation}
\rho_{l}(\varphi | x,\mathbf{P})=\delta(\varphi-\Psi_1(x,\mathbf{P}))\ \text{for $x\in
(z+x_1,z+x_2)$}\;.
\label{eq:distr_func_2}
\end{equation}

Taking into account the obtained expressions for the distribution function
(\ref{eq:rdist},\ref{eq:distr_func_1},\ref{eq:distr_func_2}), 
the integrals in (\ref{eq:int})
can be rewritten as a sum of five terms:
\begin{equation}
\begin{aligned}
&F_m(\mathbf{P})e^{iQ_m(\mathbf{P})} = \int_{\varphi_1}^{\varphi_2}d\varphi e^{im\varphi}
g\left(R(z+h)\right) \frac{\partial h}{\partial \varphi}-\\
&\int_{\varphi^b_1}^{\varphi^b_2}d\varphi e^{im\varphi}S(z+h) g\left(R(z+h)\right)
\frac{\partial h}{\partial \varphi}+
\int_{\varphi_3}^{\varphi_4}d\varphi e^{im\varphi} g\left(R(z+h)\right) \frac{\partial
h}{\partial \varphi}-\\
&\int_{\varphi^b_3}^{\varphi^b_4}d\varphi e^{im\varphi}\left(1-S(z+h)\right)
g\left(R(z+h)\right) \frac{\partial h}{\partial \varphi}
+\int_{\mathcal{X}}\int_{0}^{2\pi}dxd\varphi \frac{g(Rx)C(x)e^{im\varphi}}{|x-z-h|}\;.
\end{aligned}
\label{eq:int1}
\end{equation}
Here the first and the second terms stand for integration over the first branch $\Psi_1$ in the range $[\varphi_1,\varphi_2]$. 
The second term accounts for certain redistribution $S(x)$ of oscillators between the branches in the range $[\varphi_1^b,\varphi_2^b]$ (Fig.~\ref{fig:locked_ph}(b)).
Similarly, the third and the fourth terms correspond to integration over the possible stable branch $\Psi_2$ in the range $[\varphi_3,\varphi_4]$. 
In the same way, the fourth term accounts for redistribution of oscillators between branches in the range $[\varphi_3^b,\varphi_4^b]$ (Fig.~\ref{fig:locked_ph}(b)).
In the last term  the interval $\mathcal{X} =
(-\infty,z+H_{min}) \bigcup (z+H_{max},\infty)$  is the domain of frequencies where the
oscillators are not locked.

Now, using the 
integrals (\ref{eq:int1}), one can calculate the absolute values of the
complex
order parameters $\mathbf{X}_{1,2}$ and the frequency $\Omega$ as functions of
introduced auxiliary parameters
$R$, $u$, $v$, $z$:
\begin{equation}
X_{1,2}(\mathbf{P}) = RF_{1,2}(\mathbf{P}),\quad \theta_2 = Q_2(\mathbf{P})-2\alpha_1,\quad  \Omega(\mathbf{P}) = Rz\;.
\label{eq:self_cons1}
\end{equation}
Also, from the relations (\ref{eq:params},\ref{eq:int}) one can conclude that the following holds:
\begin{equation}
 \e_1 = \frac{\sin u}{F_1(\mathbf{P})},\quad \alpha_1 = Q_1(\mathbf{P}),\quad \gamma_1 = \frac{R\cos u}{Y},\quad \beta = \theta_y+2Q_1(\mathbf{P})-v.
\label{eq:self_cons2}
\end{equation}

\subsection{Accounting for coupling between subpopulations}
As one can see from the latter relations, the parameters of the internal 
interaction inside the first community $\e_1$ and $\alpha_1$ are determined 
only by the parameters 
$\mathbf{P}$. However, the constants of the cross-coupling $\gamma_1$ and 
$\beta$ require calculation of the order parameter $\mathbf{Y}$. 
Taking into account the transformation of variables (\ref{eq:var_tr}), 
the uniformly rotating solution of the Ott-Antonsen equation (\ref{eq:oa}) 
for the second population, the 
mean field $\mathbf{Y}$ is determined according to the following relation:
\begin{equation}
Ye^{i\theta_y}(i(2\Omega-\delta)+1)+\frac{\varepsilon_2  
Ye^{i\theta_y}}{2}(e^{-i\alpha_2}Y^2-e^{i\alpha_2}) + 
\frac{\gamma_2 X_2}{2}(e^{-i\theta_2}Y^2e^{2i\theta_y}-e^{i\theta_2})=0.
\label{eq:oa_stat}
\end{equation}
This complex equation determines $Y$ and $\theta_y$ as functions of all other parameters,
substitution of these values to Eqs.~\eqref{eq:self_cons2} will give the values of cross-coupling parameters $\gamma_1$ and 
$\beta$.

In general case solution of (\ref{eq:oa_stat}) can not be represented in an analytic form and one should use certain numerical methods to find them (a parametric representation of
solutions may be possible, but we already have four auxiliary parameters, introducing
another two appears not practical).
However, in two special cases equation (\ref{eq:oa_stat}) can be reduced to a simple polynomial equation with analytic solutions available. 
Namely, (i) for $\varepsilon_2=0$,  the problem reduces to a complex quadratic equation, 
and
(ii) for the special case $\Omega=\delta=0$ and $v=0$, equation (\ref{eq:oa_stat}) 
reduces to a real cubic equation. 
The latter case corresponds to the simplest situation when there are no phase shifts in coupling functions $\alpha_{1,2}=\beta=0$.


Summarizing, the self-consistent approach for calculation of stationary synchronous solutions of the problem (\ref{eq:opXY},\ref{eq:pheqthl}) consists of the following steps: (i) for a given set of parameters $\mathbf{P}$ and indicator function $S(x)$, one constructs the distribution function $\rho(\varphi|x)$ using microscopic dynamics (\ref{eq:bkm_first}). 
(ii) Next, using the function $\rho(\varphi|x)$ and equations (\ref{eq:int1},\ref{eq:self_cons1},\ref{eq:self_cons2}), one determines the stationary values of order parameters $\mathbf{X}_{1,2}$, rotating frequency $\Omega$ and corresponding coupling constants $\e_1$ and $\alpha_1$. 
(iii) In the following step, one should solve equation (\ref{eq:oa_stat}) for any fixed values of $\e_2$, $\alpha_2$, $\gamma_2$ and $\delta$. 
As a result, one get the stationary value for the mean field $\mathbf{Y}$ and remaining constants of the cross-coupling $\gamma_1$ and $\beta$ from (\ref{eq:self_cons2}). 

The solution is in the parametric form: varying the set of auxiliary 
parameters $\mathbf{P}$, together with $\e_2$, $\alpha_2$ and$\gamma_2$, one gets 
different solutions for the mean fields $\mathbf{X}_{1,2}$, $\mathbf{Y}$, together 
with their 
dependence on the coupling constants $\e_1$, $\gamma_1$ and $\beta$. This can be done 
for any indicator function $S(x)$, which determines re-distribution of the phases of the
first subpopulation between possible stable locked states, if multi-branch entrainment is 
possible.

In the next sections we will apply the self-consistent scheme to characterize main types of synchronous states existing in the system of two coupled subpopulations (\ref{eq:opXY},\ref{eq:pheqthl}). We will focus on the effects caused by the resonant cross coupling between the population, therefore, for the internal coupling we will consider the simplest situation when $$\alpha_{1,2}=0.$$ For the sake of simplicity, we restricts ourselves to the following parameters area: $$\e_1=\e_2=\e\ \text{and}\ \gamma_1=\gamma_2=\gamma.$$
It appears that the latter choice of parameters simplify the presentation of the results, nevertheless, it contains all the main effects peculiar for the high-order resonant interaction.

\section{Internally asynchronous populations, appearance of synchrony due to
resonant coupling}
We start with the analysis of the case when the populations are internally asynchronous, hence, 
without cross-coupling $\gamma=0$ the only stable state for each population is 
asynchrony when all mean fields vanish $X_{1,2}=0$, $Y=0$. For the 
Lorentzian distribution of frequencies, the synchronization sets in at the 
critical coupling $\e=2$. 
Therefore, in the following section we will concentrate on the case $\e<2$, i.e. the 
internal coupling inside each population is insufficient to maintain synchrony in the system
(or even repulsive).
The frequency mismatch $\delta$ together with cross-coupling constants $\gamma$ and phase shift $\beta$ constitute a set of main control parameters in the system.

\begin{figure}[!t]
\centerline{(a)\includegraphics[width=0.49\columnwidth]{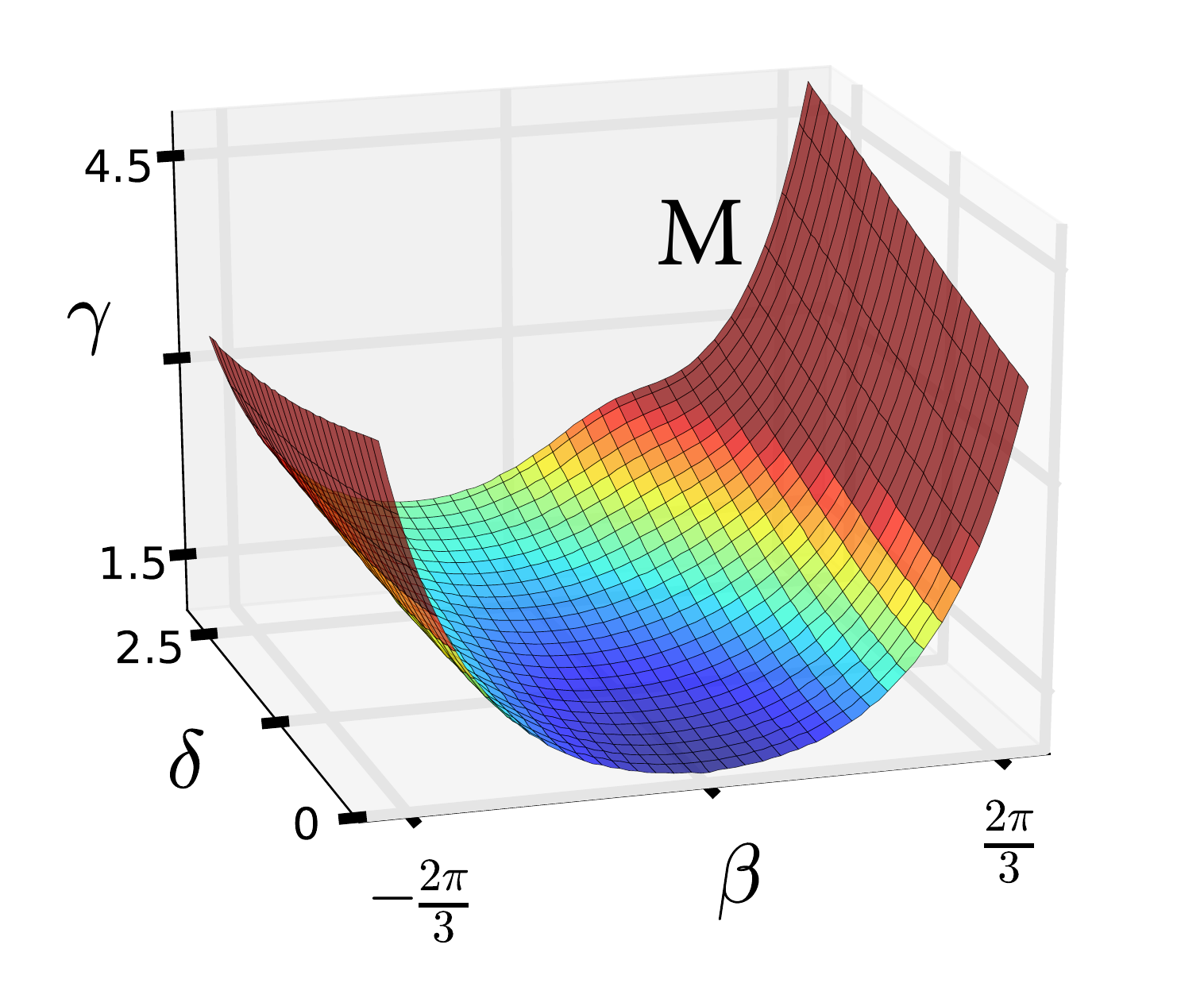}(b)\includegraphics[width=0.49\columnwidth]{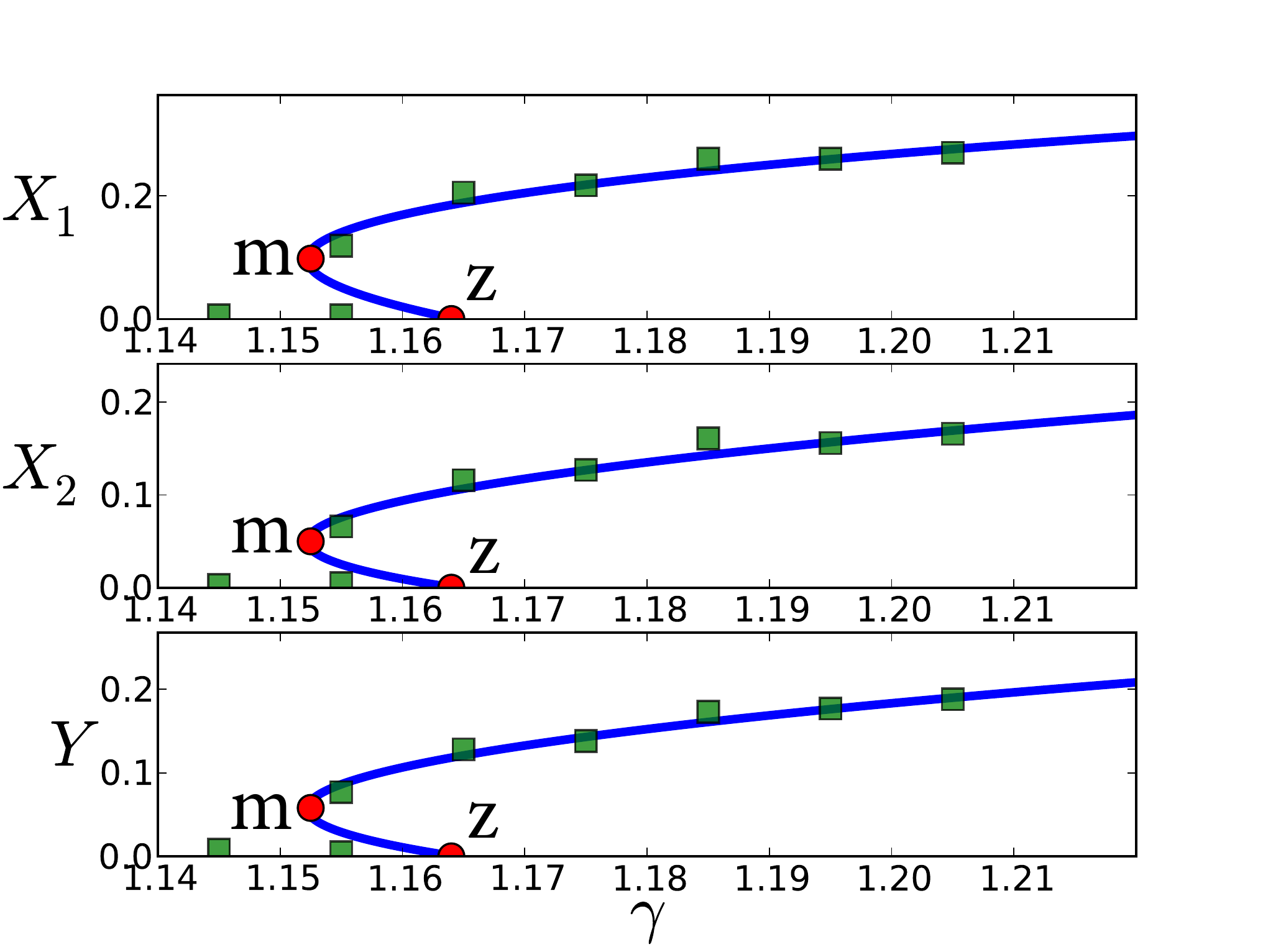}}
\centerline{(c)\includegraphics[width=0.49\columnwidth]{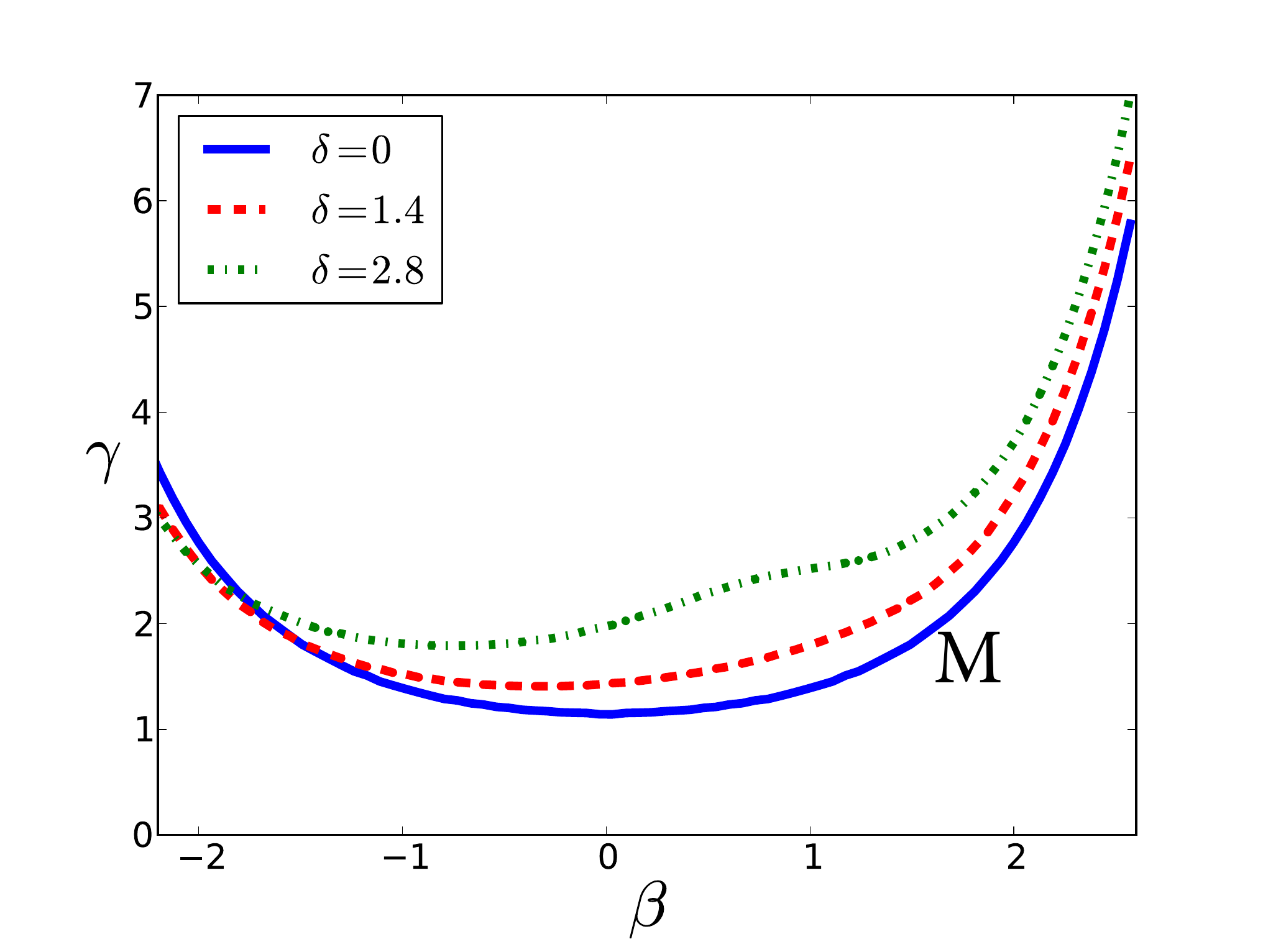} (d)\includegraphics[width=0.49\columnwidth]{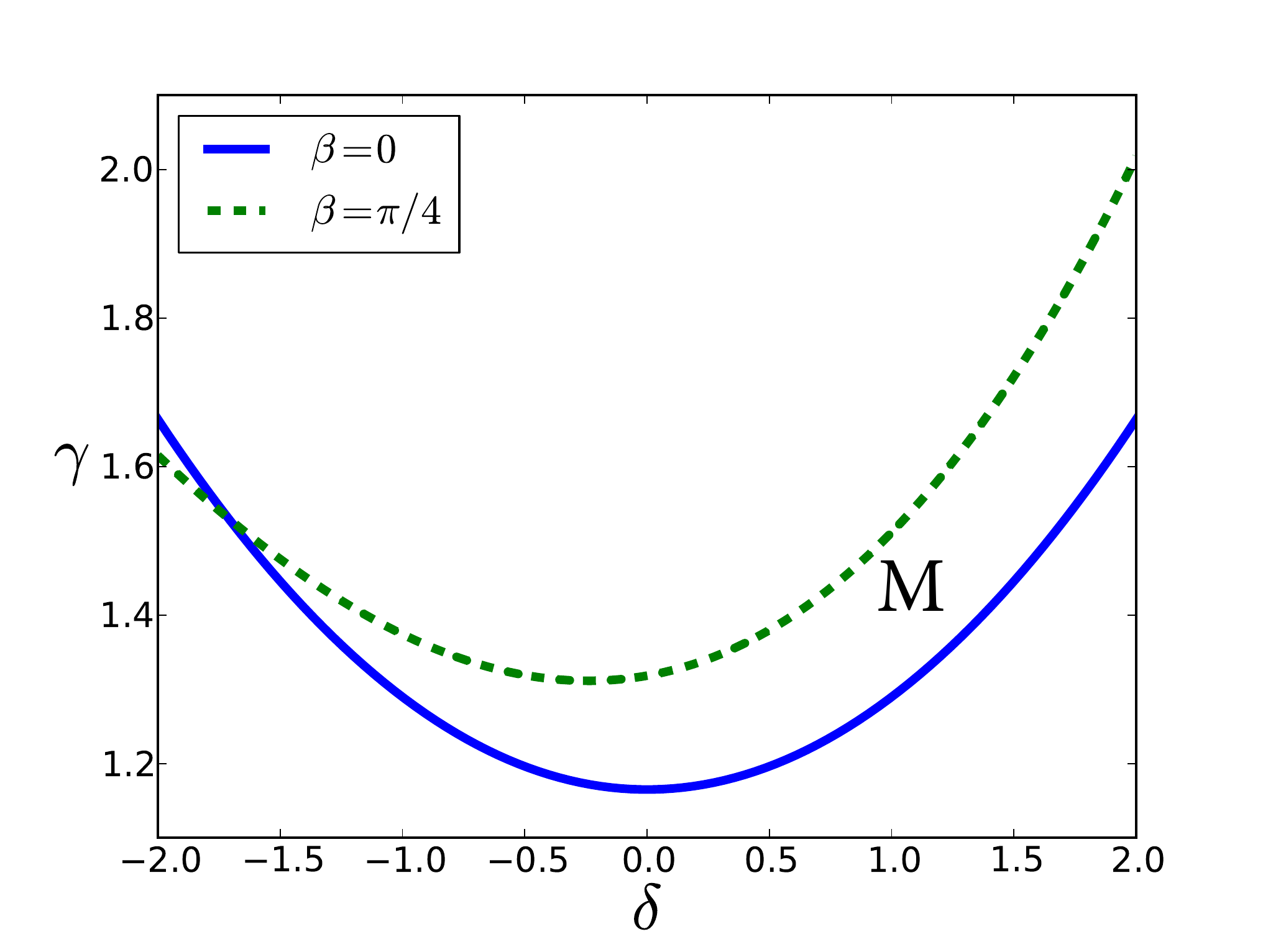}}
\caption{(a) The surface $\mathbf{M}$ depicts the boarder of synchronous states in the parameter space $(\delta,\beta,\gamma)$: above the surface synchrony with $X_{1,2}\neq 0$ and $Y\neq 0$ exists, below only asynchronous state is possible. The internal coupling $\e=1$ for each population, hence, population are internally asynchronous. (b) The dependence of order parameters on the coupling constant $\gamma$ is shown for $\delta=\beta=0$ and $\e=1$. The curves denote theoretical calculations using self-consistent scheme, markers correspond to the direct numerical calculations of the finite-size ensemble (\ref{eq:bkm_micro}) for $N=8\times 10^5$. (c,d) Cuts of the surface $\mathbf{M}$ are shown for constant values of the frequency mismatch $\delta$ (in the panel (c)) and the phase shift $\beta$ (in the panel (d)).
}
\label{fig:gamma_beta}
\end{figure}

Figure~\ref{fig:gamma_beta}(a) shows the area of existence of stationary 
synchronous solutions in the 3-d parameter space $(\delta,\beta, \gamma)$.
The surface $\mathbf{M}$ depicted in the Fig.~\ref{fig:gamma_beta}(a) denotes 
the border of existence of synchronous states: above the surface there exist stationary 
synchronous solutions with $X_{1,2}\neq 0$ and $Y\neq 0$, below $\mathbf{M}$ 
only asynchronous state exists and is stable. 
Figure~\ref{fig:gamma_beta}(b) explains the bifurcation diagram depicted 
in Fig.~\ref{fig:gamma_beta}(a), here we fix the parameters $\delta=\beta=0$ 
and plot order parameters $X_{1,2}$ and $Y$ as a function of parameter 
$\gamma$ (the latter corresponds to the vertical line passing through 
the origin in the Fig.~\ref{fig:gamma_beta}(a)). 
As one can see from the plot, there is a minimal critical coupling $\gamma_{cr}$ 
corresponding to the point $m$ where two branches of synchronous solutions arise. 
The upper branch appears to be stable, what is confirmed by direct numerical simulation of 
the finite-size ensemble. The lower branch is unstable and disappears in the point $z$, 
merging with the trivial state.
Apparently, the family of the points $m$ obtained at different values of 
$\beta$ and $\gamma$ constitutes the surface $\mathbf{M}$ depicted in 
the Fig.~\ref{fig:gamma_beta}(a). 
 
\begin{figure}[!t]
\centerline{(a)\includegraphics[width=0.49\columnwidth]{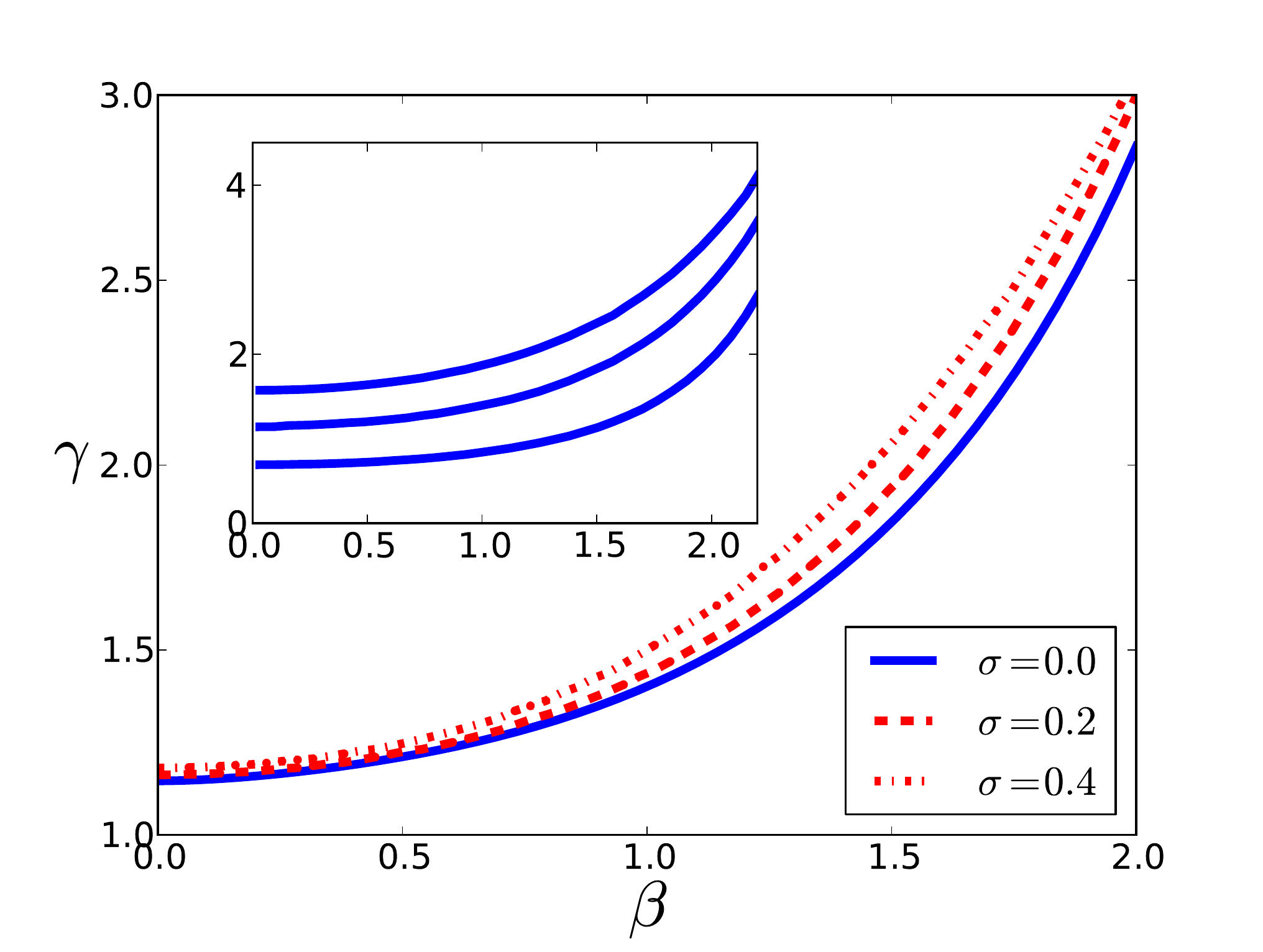}(b)\includegraphics[width=0.49\columnwidth]{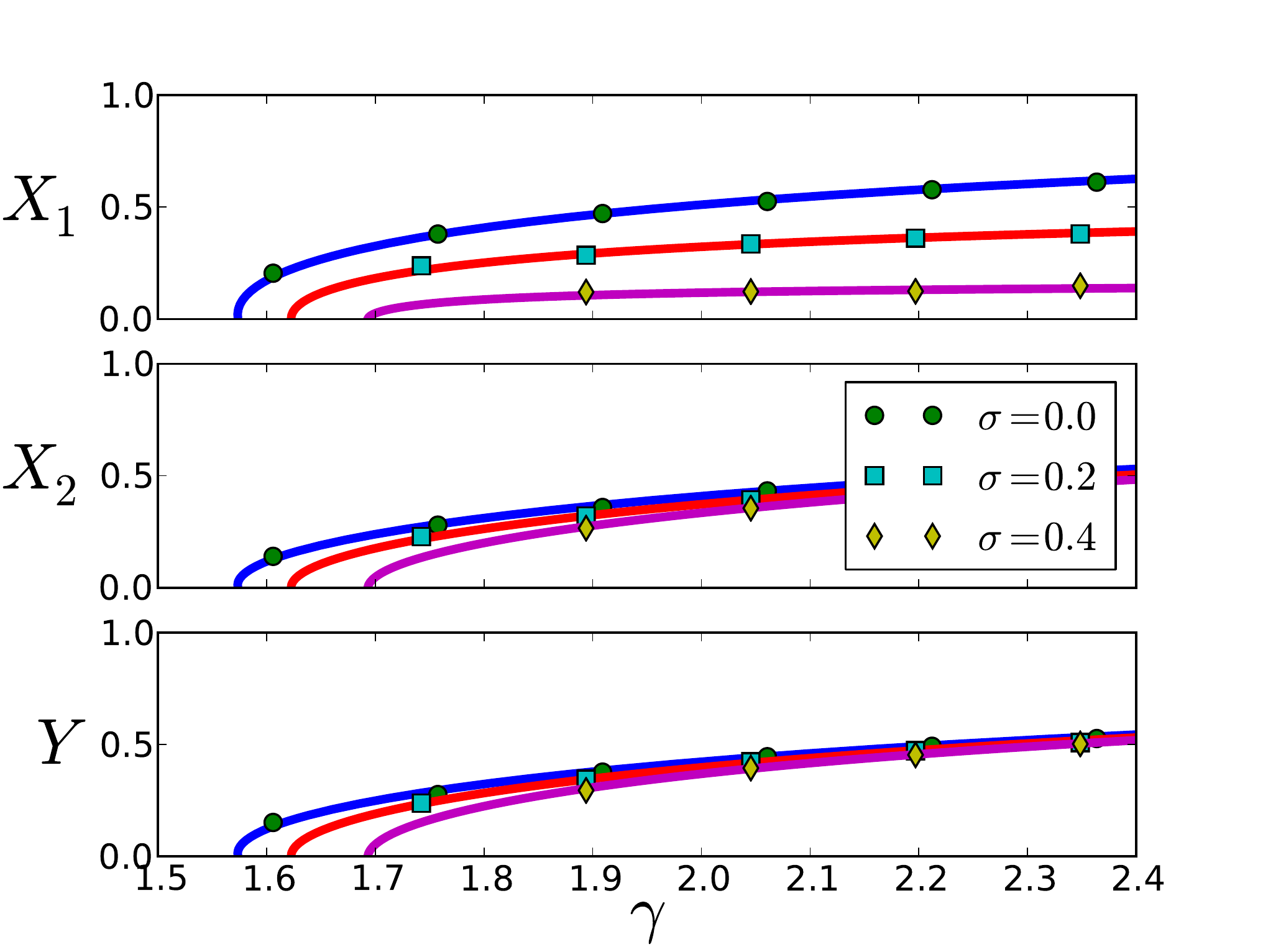} }
\centerline{(c)\includegraphics[width=0.49\columnwidth]{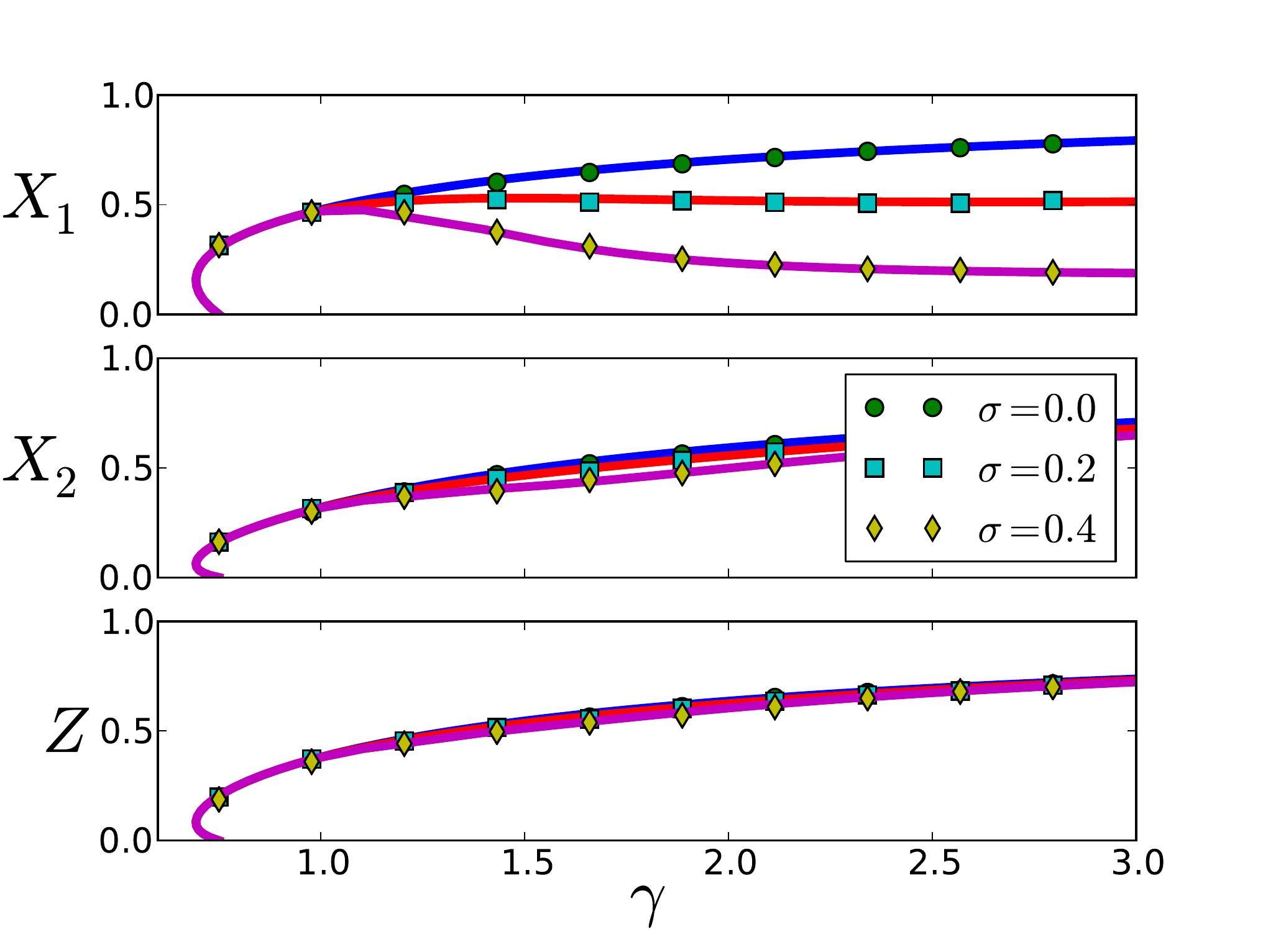} }
\caption{The areas of existence of stationary synchronous solutions on the parameter planes $(\beta,\gamma)$ are shown for the case $\e=1$ and $\delta=0$. Different curves correspond to boarder of synchronous states with different indicator functions $S(x)=\sigma=\text{const}$ (different multi-branch entrainments). Above the curves solutions with $X_{1,2}\neq 0$, $Y\neq 0$ exist. Inset shows boundaries of synchronous states plotted for different values of constant $\e$. From bottom to top $\e=1.5$, $\e=1.0$, $\e=0.5$. (b) Dependences of order parameters $X_{1,2}$, $Y$ on cross-coupling $\gamma$ are shown for states with different $\sigma$ (see legend). Solid curves denote solutions of self-consistent equations, markers denote direct calculations of the finite-size ensemble. Other parameters are: $\e=0.5$, $\beta=0$, $\delta=0$. (c) The same as (b) but for $\e=1.5$.
}
\label{fig:gamma_beta_1}
\end{figure}

The form of the surface is invariant under transformation 
$\delta \to -\delta$, $\beta\to -\beta$, that is why only 
the part with $\delta>0$ is shown in the Fig.~\ref{fig:gamma_beta}(a).
Expectedly, $\mathbf{M}$ has a global minimum at the point $\delta=\beta=0$ what means 
that, substantially, phase shift and frequency mismatch act against synchronization.  
Figures~\ref{fig:gamma_beta}(c,d) show several cuts of the surface $\mathbf{M}$ at constant 
values of $\delta=\text{const}$ (in the panel (c)) and $\beta=\text{const}$ 
(in the panel (d)).
For the most part of the parameter range, the phase shift acts against synchronization, 
as one can easily see from the Fig.~\ref{fig:gamma_beta_1}(c) where the borders 
of stationary synchronous states are plotted on the plane $(\beta,\gamma)$. 
When the frequency mismatch mismatch is absent ($\delta=0$), the curves are 
symmetric with respect to the line $\beta=0$ and critical coupling increases 
with growth of the absolute value of $\beta$. 
However, it is not always a case for non-zero frequency mismatch. 
The examples for $\delta\neq 0$ in the Fig.~\ref{fig:gamma_beta_1}(c) 
clearly indicate a nontrivial fact that the 
global minima of the curves in the $(\beta,\gamma)$ plane 
are shifted towards negative values of $\beta$.
Similarly, on the $(\delta,\gamma)$ the boarder of synchronous states has global minima at non-zero value of $\delta$ for finite phase shift $\beta=\pi/4$.

Remarkably, the transition to synchrony here is always accompanied by the 
multiplicity of different synchronous states with multi-branch 
entrainment~\citep{Daido-95,Daido-96a} in the first subpopulation.
The issue of multiplicity for the bi-harmonic Kuramoto model was studied in 
detail in~\citep{Komarov-Pikovsky-14}. 
The multiple synchronous states appear as a result of strong second 
harmonic $\sim e^{i2\varphi}$ in a global force acting on oscillators of the 
first subpopulation.
Apparently, in order to get a synchronization in the ensembles due to the cross-coupling, 
the constant $\gamma$ has to be strong enough, as one can easily see from the bifurcation 
diagram in Fig.~\ref{fig:gamma_beta}(a).
The latter implies that the coupling function $h(u,v,\varphi)$ 
(see eq.~(\ref{eq:bkm_first})) always has a 
double-well form, hence, there is always a 
possibility to redistribute oscillators between two stable branches in different ways (in 
other words, to choose an arbitrary indicator function $S(x)$ in the self-consistent scheme).
As a result, for the case of internally asynchronous populations (when $\e$ is not 
large enough), a family of synchronous states appears 
with distinct multi-branch entrainments.
Figure~\ref{fig:gamma_beta_1}(a) shows the critical couplings $\gamma$ when 
synchronous states with distinct redistributions $S(x)$ appear.
For the sake of simplicity we chose $S(x)=\sigma=\text{const}$. 
The dependences of order parameter $X_{1,2}$, $Y$ on the cross-coupling 
$\gamma$ for different types of multi-branch entrainments (characterized by 
constant $\sigma$) are presented in the Fig.~\ref{fig:gamma_beta}(b). 
The main state $\sigma=0$ arises first (i.e. at a minimal 
coupling strengths $\gamma$) in 
comparison to other states with multi-branch entrainment $\sigma\neq 0$.
Expectedly, in all cases increase of the coupling $\gamma$ leads to increase of order 
parameters values, hence, more oscillators are entrained in both populations.

Here in this section we have concentrated on the effects caused by the 
cross-coupling $\gamma$ and paid less attention to the role of the internal coupling $\e$. 
It is worth mentioning that, in the simplest form (pure sinusoidal coupling) the 
interaction inside the communities makes a relatively straightforward effect. 
Namely, increase of the coupling $\e$ leads to enlargement of the area of synchrony 
existence in the parameter space (see inset in the Fig.~\ref{fig:gamma_beta_1}).

\section{Internally synchronous populations: Chaotic dynamics.}

In this section we consider the case when $\e>2$, hence, the populations are 
internally synchronous even without cross-coupling $\gamma$.
Here we report on non-trivial phenomena when resonant cross-coupling introduces chaotic 
collective oscillations into the system.

\begin{figure}[!t]
\centerline{(a)\includegraphics[width=0.49\columnwidth]{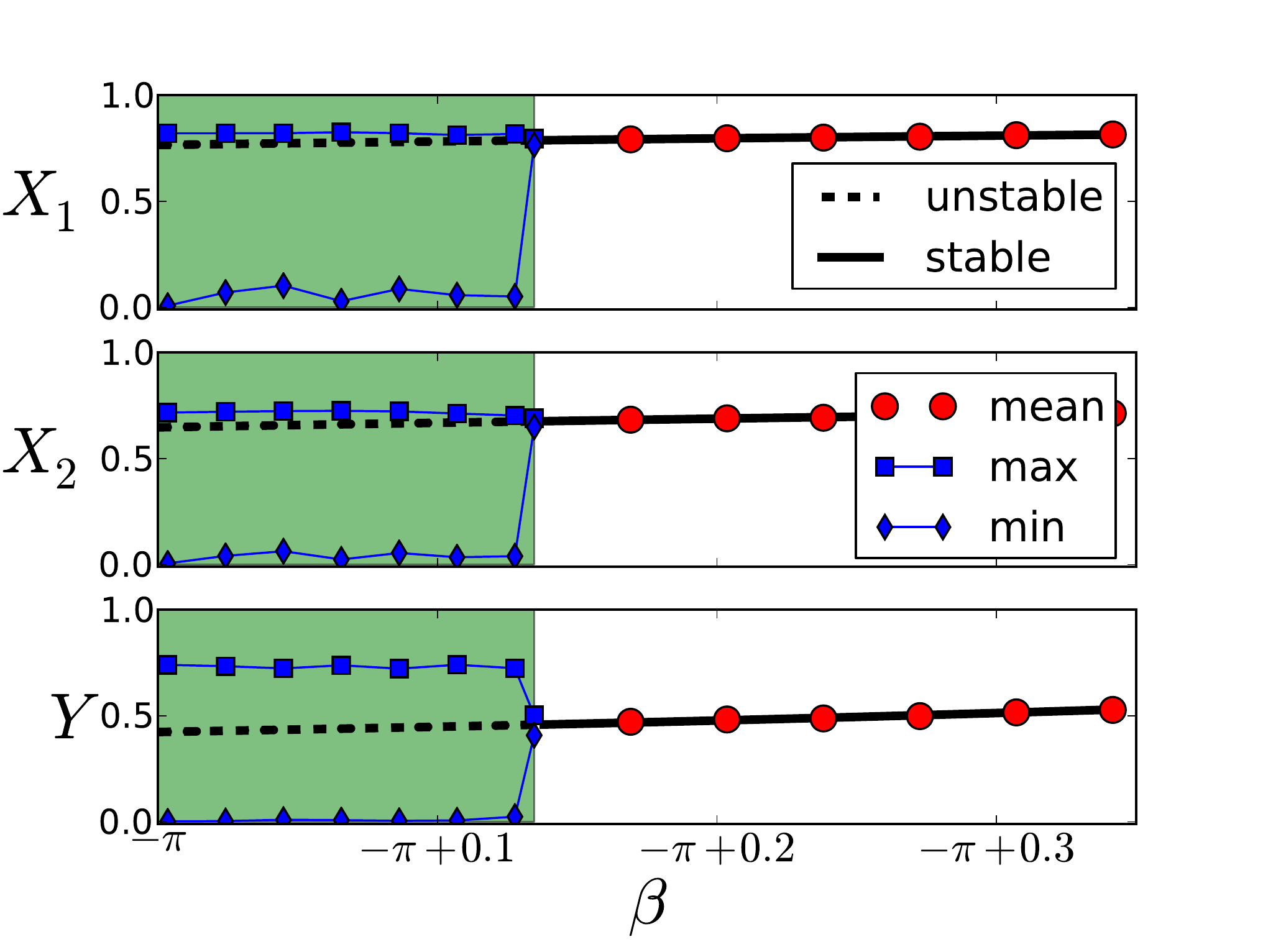}(b)\includegraphics[width=0.49\columnwidth]{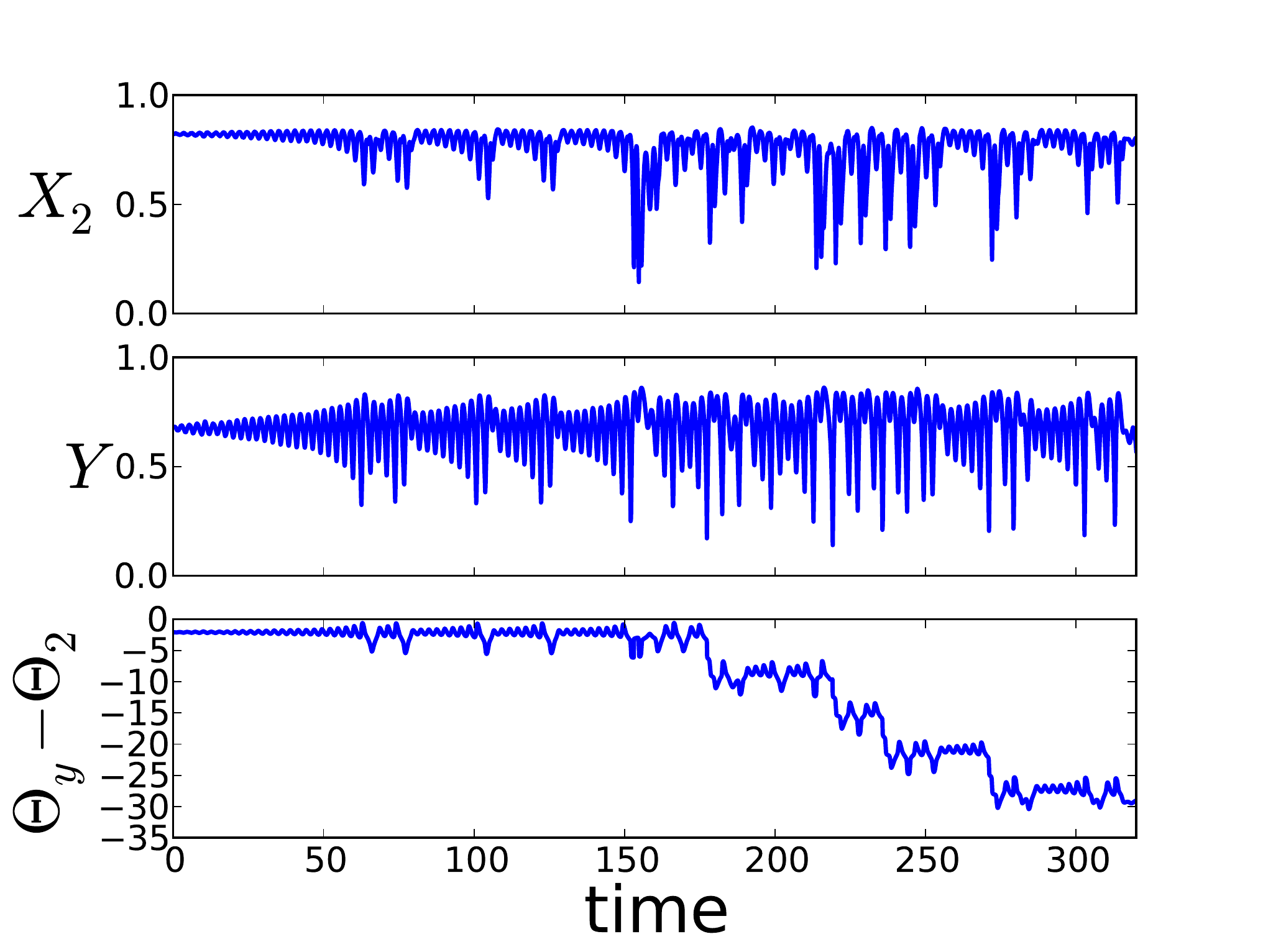}}
\centerline{(c)\includegraphics[width=0.49\columnwidth]{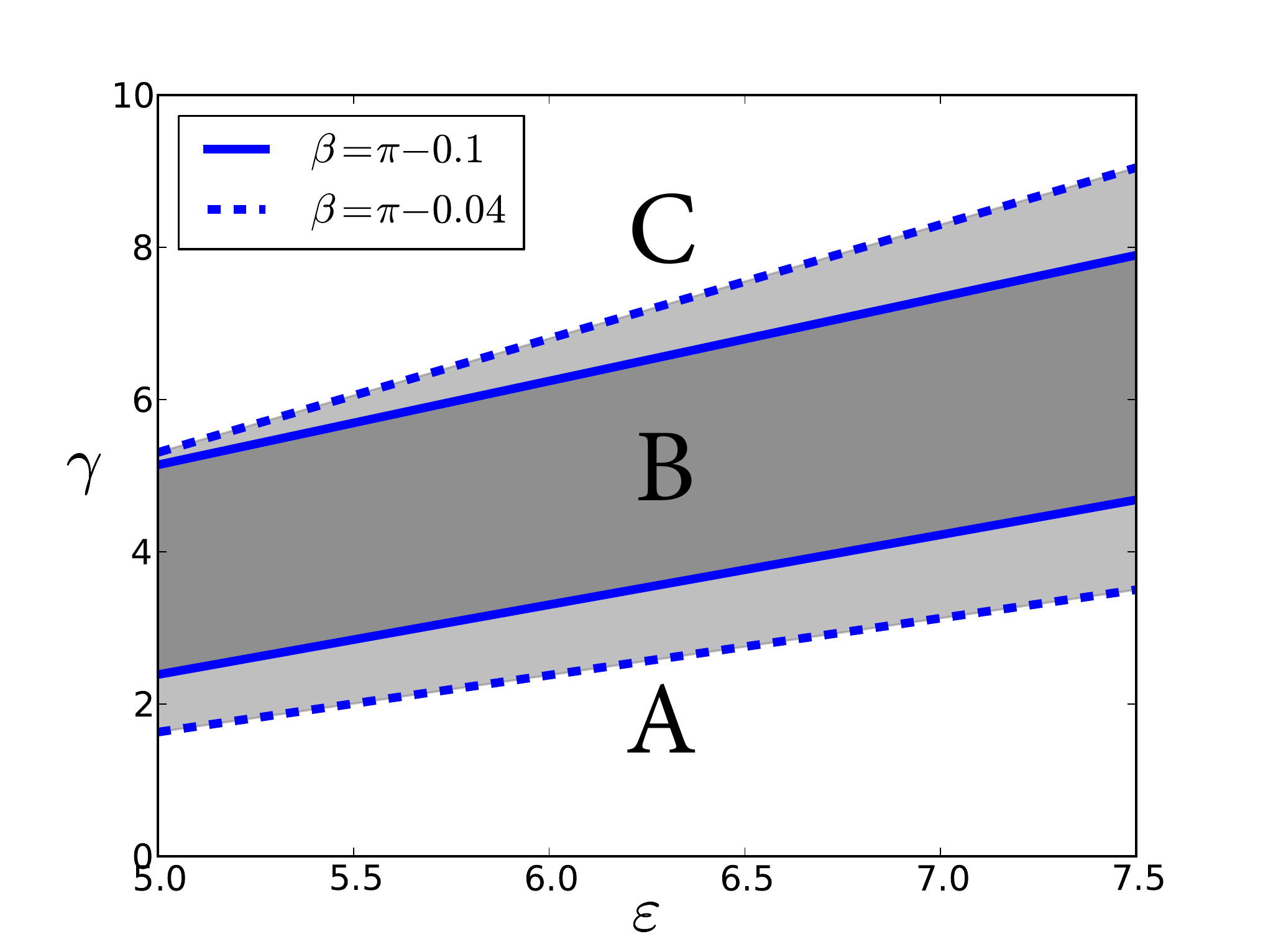}(d)\includegraphics[width=0.49\columnwidth]{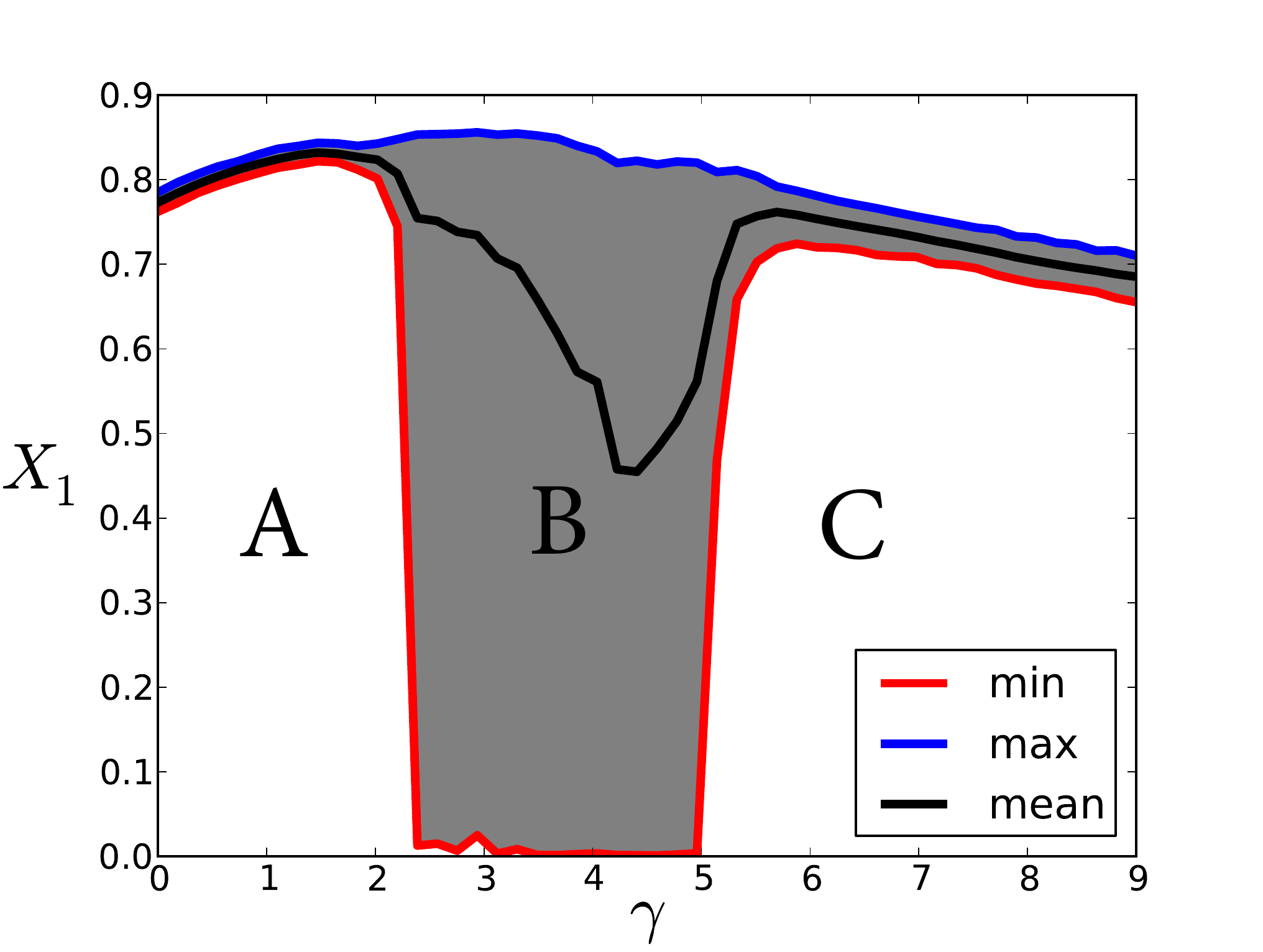}}
\caption{(a) Dependence of order parameters on phase shift $\beta$ is presented. Solid and dashed curves denote solutions obtained from the self-consistent approach. Stable solution corresponds to solid line, unstable to dashed line (stability was checked by direct simulation of the ensemble (\ref{eq:bkm_micro})). Markers correspond to simulation of finite-size ensemble (\ref{eq:bkm_micro}) for $N=10^4$ oscillators. The colored area denotes chaotic region with large amplitude of order parameters oscillations. Parameters $\e=4.5$, $\gamma=2.8$. (b) Time series of the finite-size ensembles in the chaotic regime. Parameters: $\e=4.5$, $\gamma=2.8$, $\beta = -3.0$. (c) The region of existence of chaotic mode on the parameter plane $(\varepsilon,\gamma)$ is presented for $\beta=\pi-0.1$, $\delta=0$. (d) Dependence of the order parameter $X_1$ is presented for $\varepsilon=5$, $\beta=\pi-0.1$ and $\delta=0$. The gray area denote maximal amplitude of oscillations obtained from time series after certain transient period. 
}
\label{fig:chaos}
\end{figure}

Figure~\ref{fig:chaos}(a) shows the diagram of stationary synchronous states versus 
phase shift in the cross-coupling function $\beta$. 
The solid curves correspond to solution obtained from the 
self-consistent approach above, while markers 
denote direct numerical calculations of the ensemble (\ref{eq:bkm_micro}) at the 
same parameters.
As one can easily see, the stationary states remain stable until $\beta$ is less than 
certain critical value (indicated by colored area in the Fig.~\ref{fig:chaos}(a)).
However, when the phase shift becomes relatively close to $\pi$, the synchronous solutions 
loses stability and immediately the system switches to a chaotic oscillation mode.
The corresponding time series is presented in the Fig.~\ref{fig:chaos}(b). 
Remarkably, chaotic oscillations are characterized by a drift of the phase 
difference $\Theta_2-\Theta_y$ (see the lowest panel in the Fig.~\ref{fig:chaos}(b)), so, 
the ensembles suddenly become unlocked from each other when entering to chaotic mode.

Fig.~\ref{fig:chaos}(c) is aimed to explain the 
structure of parameter area where chaotic mode exists.
As one can see, for each
sufficiently strong internal coupling $\e$, there is always a 
certain 
range of cross-coupling constant $\gamma$, where oscillations are irregular.
With an increase of the cross-coupling $\gamma$, the system pass from area $A$ (the area 
where regular synchronous solution exists) to area $B$ which corresponds to chaotic motion.
As has been mentioned above, area $B$ is characterized by a drift of the phase difference and large amplitude irregular oscillations of the order parameters (see Fig.~\ref{fig:chaos}(d)). Further increase of constant $\gamma$ leads back to a regular stationary synchronous solution (Fig.~\ref{fig:chaos}(d)). 
The size of the area $B$ on the $(\varepsilon,\gamma)$ plane is strongly related to the phase shift in coupling function: the closer the parameter $\beta$ to $\pi$, the larger the area $B$ on the $(\varepsilon,\gamma)$ plane. 

\section{Conclusions}

The phase reduction is one of the few mathematical techniques which 
allows one to perform analytical studies of complex nonlinear oscillatory systems. 
Perhaps, the most popular and well-studied phase model is the classical Kuramoto system 
which describes ensemble of globally coupled oscillators with sinusoidal type of 
interaction function. 
The derivation of various Kuramoto-type models is based on the assumption of closeness of 
natural oscillatory frequencies. 
However, in many realistic situations oscillators may have definitely different frequencies;
an example of this are neural populations that can produce brain waves wide
across the spectrum. 
For multifrequency populations one has to extend typical model 
of phase equations; in previously considered cases such an extension also
led to new dynamical regimes \citep{Komarov-Pikovsky-13,Komarov-Pikovsky-11}. 
 
 In the present paper we developed an extension of the phase synchronization theory 
 for multifrequency \textit{resonant} oscillator communities. 
 After analysing general possible resnonant terms in linear mean-field coupling, we focused 
 on the simplest high-order resonant case, 
 when two communities of 
 oscillators are globally coupled and have natural population frequencies close to the 
 rational relation 2:1.
 First, given the assumption on mean population frequencies, 
 we derived the simplest form of 
 phase equations for high-order resonant interaction between two globally coupled 
 communities of oscillators.
 Basically, the structure of the model consists of two main parts: the first part 
 represents classical sinusoidal term describing Kuramoto-type interaction inside each 
 community; the second component has different form,  it represents the resonant cross-
 coupling between the populations. 
 Next, we combine two approaches described 
 in Refs.~\cite{Komarov-Pikovsky-13a,*Komarov-Pikovsky-14} and 
 in Refs.~\cite{Ott-Antonsen-08,*Ott-Antonsen-09} to derive a self-
 consistent scheme for calculation of stationary synchronous solutions of the system 
 in  the thermodynamical limit.
 
In this paper we focused on investigation how cross-coupling promotes 
synchronization, and looked for novel dynamical effects due to the high-order resonance.
Hence, we have considered two qualitatively different cases, in the first case the 
populations were internally asynchronous, so, the internal coupling strength was relatively 
weak. Here we constructed the bifurcation diagram 
showing how synchronous regimes appear in dependence on main 
parameters of the model.
We demonstrated, that strong enough resonant cross-coupling results in stationary synchronous solutions appearing in both subpopulations. Thus, the synchrony can be only mutual. 
The nontrivial fact here is that the transition to synchrony due to the cross-coupling is 
always accompanied by multiplicity of distinct synchronous states, 
similar to the case of bi-
harmonic Kuramoto model~\cite{Komarov-Pikovsky-13a,*Komarov-Pikovsky-14}.
In the second setup, we considered an opposite situation, when the internal coupling is 
strong, such that almost all oscillators are locked to the mean fields in the absence 
of the cross-
coupling. Here we report on a quite non-trivial effect, that the cross-coupling can destroy 
the stationary synchronous state introducing chaos into the system. Mean fields of two subpopulations not only vary their amplitude chaotically, but the subpopulations also desynchronize from each other in the sense, that the phase shift between the mean 
fields is no more a constant, but performs a biased random walk.

\section{Acknowledgement}
M. K. thanks Alexander von Humboldt Foundation for support. The research in sections IV and V was supported by the Russian Science Foundation (Project 14-12-00811).


\begin{thebibliography}{45}%
\makeatletter
\providecommand \@ifxundefined [1]{%
 \@ifx{#1\undefined}
}%
\providecommand \@ifnum [1]{%
 \ifnum #1\expandafter \@firstoftwo
 \else \expandafter \@secondoftwo
 \fi
}%
\providecommand \@ifx [1]{%
 \ifx #1\expandafter \@firstoftwo
 \else \expandafter \@secondoftwo
 \fi
}%
\providecommand \natexlab [1]{#1}%
\providecommand \enquote  [1]{``#1''}%
\providecommand \bibnamefont  [1]{#1}%
\providecommand \bibfnamefont [1]{#1}%
\providecommand \citenamefont [1]{#1}%
\providecommand \href@noop [0]{\@secondoftwo}%
\providecommand \href [0]{\begingroup \@sanitize@url \@href}%
\providecommand \@href[1]{\@@startlink{#1}\@@href}%
\providecommand \@@href[1]{\endgroup#1\@@endlink}%
\providecommand \@sanitize@url [0]{\catcode `\\12\catcode `\$12\catcode
  `\&12\catcode `\#12\catcode `\^12\catcode `\_12\catcode `\%12\relax}%
\providecommand \@@startlink[1]{}%
\providecommand \@@endlink[0]{}%
\providecommand \url  [0]{\begingroup\@sanitize@url \@url }%
\providecommand \@url [1]{\endgroup\@href {#1}{\urlprefix }}%
\providecommand \urlprefix  [0]{URL }%
\providecommand \Eprint [0]{\href }%
\providecommand \doibase [0]{http://dx.doi.org/}%
\providecommand \selectlanguage [0]{\@gobble}%
\providecommand \bibinfo  [0]{\@secondoftwo}%
\providecommand \bibfield  [0]{\@secondoftwo}%
\providecommand \translation [1]{[#1]}%
\providecommand \BibitemOpen [0]{}%
\providecommand \bibitemStop [0]{}%
\providecommand \bibitemNoStop [0]{.\EOS\space}%
\providecommand \EOS [0]{\spacefactor3000\relax}%
\providecommand \BibitemShut  [1]{\csname bibitem#1\endcsname}%
\let\auto@bib@innerbib\@empty
\bibitem [{\citenamefont {Wiesenfeld}\ and\ \citenamefont
  {Swift}(1995)}]{Wiesenfeld-Swift-95}%
  \BibitemOpen
  \bibfield  {author} {\bibinfo {author} {\bibfnamefont {K.}~\bibnamefont
  {Wiesenfeld}}\ and\ \bibinfo {author} {\bibfnamefont {J.~W.}\ \bibnamefont
  {Swift}},\ }\href@noop {} {\bibfield  {journal} {\bibinfo  {journal} {Phys.
  Rev. E}\ }\textbf {\bibinfo {volume} {51}},\ \bibinfo {pages} {1020}
  (\bibinfo {year} {1995})}\BibitemShut {NoStop}%
\bibitem [{\citenamefont {Wiesenfeld}\ \emph {et~al.}(1996)\citenamefont
  {Wiesenfeld}, \citenamefont {Colet},\ and\ \citenamefont
  {Strogatz}}]{Wiesenfeld-Colet-Strogatz-96}%
  \BibitemOpen
  \bibfield  {author} {\bibinfo {author} {\bibfnamefont {K.}~\bibnamefont
  {Wiesenfeld}}, \bibinfo {author} {\bibfnamefont {P.}~\bibnamefont {Colet}}, \
  and\ \bibinfo {author} {\bibfnamefont {S.~H.}\ \bibnamefont {Strogatz}},\
  }\href@noop {} {\bibfield  {journal} {\bibinfo  {journal} {Phys. Rev. Lett.}\
  }\textbf {\bibinfo {volume} {76}},\ \bibinfo {pages} {404} (\bibinfo {year}
  {1996})}\BibitemShut {NoStop}%
\bibitem [{\citenamefont {Wiesenfeld}\ \emph {et~al.}(1998)\citenamefont
  {Wiesenfeld}, \citenamefont {Colet},\ and\ \citenamefont
  {Strogatz}}]{Wiesenfeld-Colet-Strogatz-98}%
  \BibitemOpen
  \bibfield  {author} {\bibinfo {author} {\bibfnamefont {K.}~\bibnamefont
  {Wiesenfeld}}, \bibinfo {author} {\bibfnamefont {P.}~\bibnamefont {Colet}}, \
  and\ \bibinfo {author} {\bibfnamefont {S.}~\bibnamefont {Strogatz}},\
  }\href@noop {} {\bibfield  {journal} {\bibinfo  {journal} {Physical Review
  E}\ }\textbf {\bibinfo {volume} {57}},\ \bibinfo {pages} {1563} (\bibinfo
  {year} {1998})}\BibitemShut {NoStop}%
\bibitem [{\citenamefont {Kiss}\ \emph {et~al.}(2002)\citenamefont {Kiss},
  \citenamefont {Zhai},\ and\ \citenamefont {Hudson}}]{Kiss-Zhai-Hudson-02a}%
  \BibitemOpen
  \bibfield  {author} {\bibinfo {author} {\bibfnamefont {I.}~\bibnamefont
  {Kiss}}, \bibinfo {author} {\bibfnamefont {Y.}~\bibnamefont {Zhai}}, \ and\
  \bibinfo {author} {\bibfnamefont {J.}~\bibnamefont {Hudson}},\ }\href@noop {}
  {\bibfield  {journal} {\bibinfo  {journal} {Science}\ }\textbf {\bibinfo
  {volume} {296}},\ \bibinfo {pages} {1676} (\bibinfo {year}
  {2002})}\BibitemShut {NoStop}%
\bibitem [{\citenamefont {Grollier}\ \emph {et~al.}(2006)\citenamefont
  {Grollier}, \citenamefont {Cros},\ and\ \citenamefont
  {Fert}}]{Grollier-Cros-Fert-06}%
  \BibitemOpen
  \bibfield  {author} {\bibinfo {author} {\bibfnamefont {J.}~\bibnamefont
  {Grollier}}, \bibinfo {author} {\bibfnamefont {V.}~\bibnamefont {Cros}}, \
  and\ \bibinfo {author} {\bibfnamefont {A.}~\bibnamefont {Fert}},\ }\href@noop
  {} {\bibfield  {journal} {\bibinfo  {journal} {Phys. Rev. B}\ }\textbf
  {\bibinfo {volume} {73}},\ \bibinfo {pages} {{060409(R)}} (\bibinfo {year}
  {2006})}\BibitemShut {NoStop}%
\bibitem [{\citenamefont {Georges}\ \emph {et~al.}(2008)\citenamefont
  {Georges}, \citenamefont {Grollier}, \citenamefont {Cros},\ and\
  \citenamefont {Fert}}]{georges:232504}%
  \BibitemOpen
  \bibfield  {author} {\bibinfo {author} {\bibfnamefont {B.}~\bibnamefont
  {Georges}}, \bibinfo {author} {\bibfnamefont {J.}~\bibnamefont {Grollier}},
  \bibinfo {author} {\bibfnamefont {V.}~\bibnamefont {Cros}}, \ and\ \bibinfo
  {author} {\bibfnamefont {A.}~\bibnamefont {Fert}},\ }\href@noop {} {\bibfield
   {journal} {\bibinfo  {journal} {Appl. Phys. Lett.}\ }\textbf {\bibinfo
  {volume} {92}},\ \bibinfo {eid} {232504} (\bibinfo {year}
  {2008})}\BibitemShut {NoStop}%
\bibitem [{\citenamefont {Golomb}\ \emph {et~al.}(2001)\citenamefont {Golomb},
  \citenamefont {Hansel},\ and\ \citenamefont {Mato}}]{Golomb-Hansel-Mato-01}%
  \BibitemOpen
  \bibfield  {author} {\bibinfo {author} {\bibfnamefont {D.}~\bibnamefont
  {Golomb}}, \bibinfo {author} {\bibfnamefont {D.}~\bibnamefont {Hansel}}, \
  and\ \bibinfo {author} {\bibfnamefont {G.}~\bibnamefont {Mato}},\ }in\
  \href@noop {} {\emph {\bibinfo {booktitle} {{Neuro-informatics and Neural
  Modeling}}}},\ \bibinfo {series} {{Handbook of Biological Physics}},
  Vol.~\bibinfo {volume} {4},\ \bibinfo {editor} {edited by\ \bibinfo {editor}
  {\bibfnamefont {F.}~\bibnamefont {Moss}}\ and\ \bibinfo {editor}
  {\bibfnamefont {S.}~\bibnamefont {Gielen}}}\ (\bibinfo  {publisher}
  {Elsevier},\ \bibinfo {address} {Amsterdam},\ \bibinfo {year} {2001})\ pp.\
  \bibinfo {pages} {887--968}\BibitemShut {NoStop}%
\bibitem [{\citenamefont {Breakspear}\ \emph {et~al.}(2010)\citenamefont
  {Breakspear}, \citenamefont {Heitmann},\ and\ \citenamefont
  {Daffertshofer}}]{Breakspear-Heitmann-Daffertshofer-10}%
  \BibitemOpen
  \bibfield  {author} {\bibinfo {author} {\bibfnamefont {M.}~\bibnamefont
  {Breakspear}}, \bibinfo {author} {\bibfnamefont {S.}~\bibnamefont
  {Heitmann}}, \ and\ \bibinfo {author} {\bibfnamefont {A.}~\bibnamefont
  {Daffertshofer}},\ }\href@noop {} {\bibfield  {journal} {\bibinfo  {journal}
  {{Frontiers in human neuroscience}}\ }\textbf {\bibinfo {volume} {{4}}},\
  \bibinfo {pages} {{190}} (\bibinfo {year} {{2010}})}\BibitemShut {NoStop}%
\bibitem [{\citenamefont {Gonze}\ \emph {et~al.}(2005)\citenamefont {Gonze},
  \citenamefont {Bernard}, \citenamefont {Waltermann}, \citenamefont {Kramer},\
  and\ \citenamefont {Herzel}}]{Gonze-05}%
  \BibitemOpen
  \bibfield  {author} {\bibinfo {author} {\bibfnamefont {D.}~\bibnamefont
  {Gonze}}, \bibinfo {author} {\bibfnamefont {S.}~\bibnamefont {Bernard}},
  \bibinfo {author} {\bibfnamefont {C.}~\bibnamefont {Waltermann}}, \bibinfo
  {author} {\bibfnamefont {A.}~\bibnamefont {Kramer}}, \ and\ \bibinfo {author}
  {\bibfnamefont {H.}~\bibnamefont {Herzel}},\ }\href@noop {} {\bibfield
  {journal} {\bibinfo  {journal} {Biophysical Journal}\ }\textbf {\bibinfo
  {volume} {89}},\ \bibinfo {pages} {120 } (\bibinfo {year}
  {2005})}\BibitemShut {NoStop}%
\bibitem [{\citenamefont {Bordyugov}\ \emph {et~al.}(2013)\citenamefont
  {Bordyugov}, \citenamefont {Westermark}, \citenamefont {Korencic},\ and\
  \citenamefont {Herzel}}]{Bordyugov-13}%
  \BibitemOpen
  \bibfield  {author} {\bibinfo {author} {\bibfnamefont {G.}~\bibnamefont
  {Bordyugov}}, \bibinfo {author} {\bibfnamefont {P.}~\bibnamefont
  {Westermark}}, \bibinfo {author} {\bibfnamefont {A.}~\bibnamefont
  {Korencic}}, \ and\ \bibinfo {author} {\bibfnamefont {H.}~\bibnamefont
  {Herzel}},\ }in\ \href@noop {} {\emph {\bibinfo {booktitle} {{Circadian
  Clocks}}}},\ \bibinfo {editor} {edited by\ \bibinfo {editor} {\bibfnamefont
  {A.}~\bibnamefont {Kramer}}\ and\ \bibinfo {editor} {\bibfnamefont
  {M.}~\bibnamefont {Merrow}}}\ (\bibinfo  {publisher} {Springer},\ \bibinfo
  {year} {2013})\BibitemShut {NoStop}%
\bibitem [{\citenamefont {Eckhardt}\ \emph {et~al.}(2007)\citenamefont
  {Eckhardt}, \citenamefont {Ott}, \citenamefont {Strogatz}, \citenamefont
  {Abrams},\ and\ \citenamefont {McRobie}}]{Eckhardt_et_al-07}%
  \BibitemOpen
  \bibfield  {author} {\bibinfo {author} {\bibfnamefont {B.}~\bibnamefont
  {Eckhardt}}, \bibinfo {author} {\bibfnamefont {E.}~\bibnamefont {Ott}},
  \bibinfo {author} {\bibfnamefont {S.~H.}\ \bibnamefont {Strogatz}}, \bibinfo
  {author} {\bibfnamefont {D.~M.}\ \bibnamefont {Abrams}}, \ and\ \bibinfo
  {author} {\bibfnamefont {A.}~\bibnamefont {McRobie}},\ }\href@noop {}
  {\bibfield  {journal} {\bibinfo  {journal} {Phys. Rev. E}\ }\textbf {\bibinfo
  {volume} {75}},\ \bibinfo {pages} {021110} (\bibinfo {year}
  {2007})}\BibitemShut {NoStop}%
\bibitem [{\citenamefont {N\'eda}\ \emph {et~al.}(2000)\citenamefont {N\'eda},
  \citenamefont {Ravasz}, \citenamefont {Vicsek}, \citenamefont {Brechet},\
  and\ \citenamefont {Barab\'asi}}]{Neda_etal-00}%
  \BibitemOpen
  \bibfield  {author} {\bibinfo {author} {\bibfnamefont {Z.}~\bibnamefont
  {N\'eda}}, \bibinfo {author} {\bibfnamefont {E.}~\bibnamefont {Ravasz}},
  \bibinfo {author} {\bibfnamefont {T.}~\bibnamefont {Vicsek}}, \bibinfo
  {author} {\bibfnamefont {Y.}~\bibnamefont {Brechet}}, \ and\ \bibinfo
  {author} {\bibfnamefont {A.~L.}\ \bibnamefont {Barab\'asi}},\ }\href@noop {}
  {\bibfield  {journal} {\bibinfo  {journal} {Phys. Rev. E}\ }\textbf {\bibinfo
  {volume} {61}},\ \bibinfo {pages} {6987} (\bibinfo {year}
  {2000})}\BibitemShut {NoStop}%
\bibitem [{\citenamefont {Kuramoto}(1984)}]{Kuramoto-84}%
  \BibitemOpen
  \bibfield  {author} {\bibinfo {author} {\bibfnamefont {Y.}~\bibnamefont
  {Kuramoto}},\ }\href@noop {} {\emph {\bibinfo {title} {Chemical Oscillations,
  Waves and Turbulence}}}\ (\bibinfo  {publisher} {Springer},\ \bibinfo
  {address} {Berlin},\ \bibinfo {year} {1984})\BibitemShut {NoStop}%
\bibitem [{\citenamefont {Pikovsky}\ \emph {et~al.}(2001)\citenamefont
  {Pikovsky}, \citenamefont {Rosenblum},\ and\ \citenamefont
  {Kurths}}]{Pikovsky-Rosenblum-Kurths-01}%
  \BibitemOpen
  \bibfield  {author} {\bibinfo {author} {\bibfnamefont {A.}~\bibnamefont
  {Pikovsky}}, \bibinfo {author} {\bibfnamefont {M.}~\bibnamefont {Rosenblum}},
  \ and\ \bibinfo {author} {\bibfnamefont {J.}~\bibnamefont {Kurths}},\
  }\href@noop {} {\emph {\bibinfo {title} {Synchronization. A Universal Concept
  in Nonlinear Sciences.}}}\ (\bibinfo  {publisher} {Cambridge University
  Press},\ \bibinfo {address} {Cambridge},\ \bibinfo {year} {2001})\BibitemShut
  {NoStop}%
\bibitem [{\citenamefont {Kuramoto}(1975)}]{Kuramoto-75}%
  \BibitemOpen
  \bibfield  {author} {\bibinfo {author} {\bibfnamefont {Y.}~\bibnamefont
  {Kuramoto}},\ }in\ \href@noop {} {\emph {\bibinfo {booktitle} {International
  Symposium on Mathematical Problems in Theoretical Physics}}},\ \bibinfo
  {editor} {edited by\ \bibinfo {editor} {\bibfnamefont {H.}~\bibnamefont
  {Araki}}}\ (\bibinfo  {publisher} {Springer Lecture Notes Phys., v. 39},\
  \bibinfo {address} {New York},\ \bibinfo {year} {1975})\ p.\ \bibinfo {pages}
  {420}\BibitemShut {NoStop}%
\bibitem [{\citenamefont {Izhikevich}(2007)}]{Izhikevich-07}%
  \BibitemOpen
  \bibfield  {author} {\bibinfo {author} {\bibfnamefont {E.~M.}\ \bibnamefont
  {Izhikevich}},\ }\href@noop {} {\emph {\bibinfo {title} {Dynamical Systems in
  Neuroscience}}}\ (\bibinfo  {publisher} {MIT Press},\ \bibinfo {address}
  {Cambridge, Mass.},\ \bibinfo {year} {2007})\BibitemShut {NoStop}%
\bibitem [{\citenamefont {Izhikevich}(2000)}]{Izhikevich-00}%
  \BibitemOpen
  \bibfield  {author} {\bibinfo {author} {\bibfnamefont {E.~M.}\ \bibnamefont
  {Izhikevich}},\ }\href@noop {} {\bibfield  {journal} {\bibinfo  {journal}
  {SIAM Journal on Applied Mathematics}\ }\textbf {\bibinfo {volume} {60}},\
  \bibinfo {pages} {1789} (\bibinfo {year} {2000})}\BibitemShut {NoStop}%
\bibitem [{\citenamefont {Daido}(1993{\natexlab{a}})}]{Daido-93}%
  \BibitemOpen
  \bibfield  {author} {\bibinfo {author} {\bibfnamefont {H.}~\bibnamefont
  {Daido}},\ }\href@noop {} {\bibfield  {journal} {\bibinfo  {journal} {Physica
  D}\ }\textbf {\bibinfo {volume} {69}},\ \bibinfo {pages} {394} (\bibinfo
  {year} {1993}{\natexlab{a}})}\BibitemShut {NoStop}%
\bibitem [{\citenamefont {Daido}(1993{\natexlab{b}})}]{Daido-93a}%
  \BibitemOpen
  \bibfield  {author} {\bibinfo {author} {\bibfnamefont {H.}~\bibnamefont
  {Daido}},\ }\href@noop {} {\bibfield  {journal} {\bibinfo  {journal} {Prog.
  Theor. Phys.}\ }\textbf {\bibinfo {volume} {89}},\ \bibinfo {pages} {929}
  (\bibinfo {year} {1993}{\natexlab{b}})}\BibitemShut {NoStop}%
\bibitem [{\citenamefont {Daido}(1996{\natexlab{a}})}]{Daido-96}%
  \BibitemOpen
  \bibfield  {author} {\bibinfo {author} {\bibfnamefont {H.}~\bibnamefont
  {Daido}},\ }\href@noop {} {\bibfield  {journal} {\bibinfo  {journal} {Physica
  D}\ }\textbf {\bibinfo {volume} {91}},\ \bibinfo {pages} {24} (\bibinfo
  {year} {1996}{\natexlab{a}})}\BibitemShut {NoStop}%
\bibitem [{\citenamefont {Daido}(1995)}]{Daido-95}%
  \BibitemOpen
  \bibfield  {author} {\bibinfo {author} {\bibfnamefont {H.}~\bibnamefont
  {Daido}},\ }\href@noop {} {\bibfield  {journal} {\bibinfo  {journal} {J.
  Phys. A: Math. Gen.}\ }\textbf {\bibinfo {volume} {28}},\ \bibinfo {pages}
  {L151} (\bibinfo {year} {1995})}\BibitemShut {NoStop}%
\bibitem [{\citenamefont {Acebr{\'o}n}\ \emph {et~al.}(2005)\citenamefont
  {Acebr{\'o}n}, \citenamefont {Bonilla}, \citenamefont {Vicente},
  \citenamefont {Ritort},\ and\ \citenamefont {Spigler}}]{Acebron-etal-05}%
  \BibitemOpen
  \bibfield  {author} {\bibinfo {author} {\bibfnamefont {J.~A.}\ \bibnamefont
  {Acebr{\'o}n}}, \bibinfo {author} {\bibfnamefont {L.~L.}\ \bibnamefont
  {Bonilla}}, \bibinfo {author} {\bibfnamefont {C.~J.~P.}\ \bibnamefont
  {Vicente}}, \bibinfo {author} {\bibfnamefont {F.}~\bibnamefont {Ritort}}, \
  and\ \bibinfo {author} {\bibfnamefont {R.}~\bibnamefont {Spigler}},\
  }\href@noop {} {\bibfield  {journal} {\bibinfo  {journal} {Rev. Mod. Phys.}\
  }\textbf {\bibinfo {volume} {77}},\ \bibinfo {pages} {137} (\bibinfo {year}
  {2005})}\BibitemShut {NoStop}%
\bibitem [{\citenamefont {Watanabe}\ and\ \citenamefont
  {Strogatz}(1994)}]{Watanabe-Strogatz-94}%
  \BibitemOpen
  \bibfield  {author} {\bibinfo {author} {\bibfnamefont {S.}~\bibnamefont
  {Watanabe}}\ and\ \bibinfo {author} {\bibfnamefont {S.~H.}\ \bibnamefont
  {Strogatz}},\ }\href@noop {} {\bibfield  {journal} {\bibinfo  {journal}
  {Physica D}\ }\textbf {\bibinfo {volume} {74}},\ \bibinfo {pages} {197}
  (\bibinfo {year} {1994})}\BibitemShut {NoStop}%
\bibitem [{\citenamefont {Ott}\ and\ \citenamefont
  {Antonsen}(2008)}]{Ott-Antonsen-08}%
  \BibitemOpen
  \bibfield  {author} {\bibinfo {author} {\bibfnamefont {E.}~\bibnamefont
  {Ott}}\ and\ \bibinfo {author} {\bibfnamefont {T.~M.}\ \bibnamefont
  {Antonsen}},\ }\href@noop {} {\bibfield  {journal} {\bibinfo  {journal}
  {CHAOS}\ }\textbf {\bibinfo {volume} {18}},\ \bibinfo {pages} {037113}
  (\bibinfo {year} {2008})}\BibitemShut {NoStop}%
\bibitem [{\citenamefont {Ott}\ and\ \citenamefont
  {Antonsen}(2009)}]{Ott-Antonsen-09}%
  \BibitemOpen
  \bibfield  {author} {\bibinfo {author} {\bibfnamefont {E.}~\bibnamefont
  {Ott}}\ and\ \bibinfo {author} {\bibfnamefont {T.~M.}\ \bibnamefont
  {Antonsen}},\ }\href@noop {} {\bibfield  {journal} {\bibinfo  {journal}
  {{CHAOS}}\ }\textbf {\bibinfo {volume} {{19}}},\ \bibinfo {pages} {{023117}}
  (\bibinfo {year} {{2009}})}\BibitemShut {NoStop}%
\bibitem [{\citenamefont {Marvel}\ \emph {et~al.}(2009)\citenamefont {Marvel},
  \citenamefont {Mirollo},\ and\ \citenamefont
  {Strogatz}}]{Marvel-Mirollo-Strogatz-09}%
  \BibitemOpen
  \bibfield  {author} {\bibinfo {author} {\bibfnamefont {S.~A.}\ \bibnamefont
  {Marvel}}, \bibinfo {author} {\bibfnamefont {R.~E.}\ \bibnamefont {Mirollo}},
  \ and\ \bibinfo {author} {\bibfnamefont {S.~H.}\ \bibnamefont {Strogatz}},\
  }\href@noop {} {\bibfield  {journal} {\bibinfo  {journal} {Chaos}\ }\textbf
  {\bibinfo {volume} {19}},\ \bibinfo {pages} {043104.} (\bibinfo {year}
  {2009})}\BibitemShut {NoStop}%
\bibitem [{\citenamefont {Pikovsky}\ and\ \citenamefont
  {Rosenblum}(2008)}]{Pikovsky-Rosenblum-08}%
  \BibitemOpen
  \bibfield  {author} {\bibinfo {author} {\bibfnamefont {A.}~\bibnamefont
  {Pikovsky}}\ and\ \bibinfo {author} {\bibfnamefont {M.}~\bibnamefont
  {Rosenblum}},\ }\href@noop {} {\bibfield  {journal} {\bibinfo  {journal}
  {Phys. Rev. Lett.}\ }\textbf {\bibinfo {volume} {101}},\ \bibinfo {pages}
  {264103} (\bibinfo {year} {2008})}\BibitemShut {NoStop}%
\bibitem [{\citenamefont {Pikovsky}\ and\ \citenamefont
  {Rosenblum}(2011)}]{Pikovsky-Rosenblum-11}%
  \BibitemOpen
  \bibfield  {author} {\bibinfo {author} {\bibfnamefont {A.}~\bibnamefont
  {Pikovsky}}\ and\ \bibinfo {author} {\bibfnamefont {M.}~\bibnamefont
  {Rosenblum}},\ }\href@noop {} {\bibfield  {journal} {\bibinfo  {journal}
  {Physica D}\ }\textbf {\bibinfo {volume} {240}},\ \bibinfo {pages} {872}
  (\bibinfo {year} {2011})}\BibitemShut {NoStop}%
\bibitem [{\citenamefont {Omel'chenko}\ and\ \citenamefont
  {Wolfrum}(2012)}]{Omelchenko-Wolfrum-12}%
  \BibitemOpen
  \bibfield  {author} {\bibinfo {author} {\bibfnamefont {O.~E.}\ \bibnamefont
  {Omel'chenko}}\ and\ \bibinfo {author} {\bibfnamefont {M.}~\bibnamefont
  {Wolfrum}},\ }\href@noop {} {\bibfield  {journal} {\bibinfo  {journal} {Phys.
  Rev. Lett.}\ }\textbf {\bibinfo {volume} {109}},\ \bibinfo {pages} {164101}
  (\bibinfo {year} {2012})}\BibitemShut {NoStop}%
\bibitem [{\citenamefont {Omel'chenko}\ and\ \citenamefont
  {Wolfrum}(2013)}]{Omelchenko-Wolfrum-13}%
  \BibitemOpen
  \bibfield  {author} {\bibinfo {author} {\bibfnamefont {O.~E.}\ \bibnamefont
  {Omel'chenko}}\ and\ \bibinfo {author} {\bibfnamefont {M.}~\bibnamefont
  {Wolfrum}},\ }\href@noop {} {\bibfield  {journal} {\bibinfo  {journal}
  {Physica D}\ }\textbf {\bibinfo {volume} {263}},\ \bibinfo {pages} {74}
  (\bibinfo {year} {2013})}\BibitemShut {NoStop}%
\bibitem [{\citenamefont {Daido}(1996{\natexlab{b}})}]{Daido-96a}%
  \BibitemOpen
  \bibfield  {author} {\bibinfo {author} {\bibfnamefont {H.}~\bibnamefont
  {Daido}},\ }\href@noop {} {\bibfield  {journal} {\bibinfo  {journal} {Phys.
  Rev. Lett.}\ }\textbf {\bibinfo {volume} {77}},\ \bibinfo {pages} {1406}
  (\bibinfo {year} {1996}{\natexlab{b}})}\BibitemShut {NoStop}%
\bibitem [{\citenamefont {Crawford}(1995)}]{Crawford-95}%
  \BibitemOpen
  \bibfield  {author} {\bibinfo {author} {\bibfnamefont {J.~D.}\ \bibnamefont
  {Crawford}},\ }\href@noop {} {\bibfield  {journal} {\bibinfo  {journal}
  {Phys. Rev. Lett.}\ }\textbf {\bibinfo {volume} {74}},\ \bibinfo {pages}
  {4341} (\bibinfo {year} {1995})}\BibitemShut {NoStop}%
\bibitem [{\citenamefont {Crawford}\ and\ \citenamefont
  {Davies}(1999)}]{Crawford-Davies-99}%
  \BibitemOpen
  \bibfield  {author} {\bibinfo {author} {\bibfnamefont {J.~D.}\ \bibnamefont
  {Crawford}}\ and\ \bibinfo {author} {\bibfnamefont {K.~T.~R.}\ \bibnamefont
  {Davies}},\ }\href@noop {} {\bibfield  {journal} {\bibinfo  {journal}
  {Physica D}\ }\textbf {\bibinfo {volume} {125}},\ \bibinfo {pages} {1}
  (\bibinfo {year} {1999})}\BibitemShut {NoStop}%
\bibitem [{\citenamefont {Chiba}\ and\ \citenamefont
  {Nishikawa}(2011)}]{Chiba-Nishikawa-11}%
  \BibitemOpen
  \bibfield  {author} {\bibinfo {author} {\bibfnamefont {H.}~\bibnamefont
  {Chiba}}\ and\ \bibinfo {author} {\bibfnamefont {I.}~\bibnamefont
  {Nishikawa}},\ }\href@noop {} {\bibfield  {journal} {\bibinfo  {journal}
  {Chaos}\ }\textbf {\bibinfo {volume} {21}},\ \bibinfo {eid} {043103}
  (\bibinfo {year} {2011})}\BibitemShut {NoStop}%
\bibitem [{\citenamefont {Hansel}\ \emph {et~al.}(1993)\citenamefont {Hansel},
  \citenamefont {Mato},\ and\ \citenamefont {Meunier}}]{Hansel-93}%
  \BibitemOpen
  \bibfield  {author} {\bibinfo {author} {\bibfnamefont {D.}~\bibnamefont
  {Hansel}}, \bibinfo {author} {\bibfnamefont {G.}~\bibnamefont {Mato}}, \ and\
  \bibinfo {author} {\bibfnamefont {C.}~\bibnamefont {Meunier}},\ }\href@noop
  {} {\bibfield  {journal} {\bibinfo  {journal} {Phys. Rev. E}\ }\textbf
  {\bibinfo {volume} {48}},\ \bibinfo {pages} {3470} (\bibinfo {year}
  {1993})}\BibitemShut {NoStop}%
\bibitem [{\citenamefont {Ashwin}\ \emph {et~al.}(2007)\citenamefont {Ashwin},
  \citenamefont {Orosz}, \citenamefont {Wordsworth},\ and\ \citenamefont
  {Townley}}]{Ashwin_etal-07}%
  \BibitemOpen
  \bibfield  {author} {\bibinfo {author} {\bibfnamefont {P.}~\bibnamefont
  {Ashwin}}, \bibinfo {author} {\bibfnamefont {G.}~\bibnamefont {Orosz}},
  \bibinfo {author} {\bibfnamefont {J.}~\bibnamefont {Wordsworth}}, \ and\
  \bibinfo {author} {\bibfnamefont {S.}~\bibnamefont {Townley}},\ }\href
  {\doibase 10.1137/070683969} {\bibfield  {journal} {\bibinfo  {journal} {SIAM
  Journal on Applied Dynamical Systems}\ }\textbf {\bibinfo {volume} {6}},\
  \bibinfo {pages} {728} (\bibinfo {year} {2007})}\BibitemShut {NoStop}%
\bibitem [{\citenamefont {Komarov}\ and\ \citenamefont
  {Pikovsky}(2013{\natexlab{a}})}]{Komarov-Pikovsky-13a}%
  \BibitemOpen
  \bibfield  {author} {\bibinfo {author} {\bibfnamefont {M.}~\bibnamefont
  {Komarov}}\ and\ \bibinfo {author} {\bibfnamefont {A.}~\bibnamefont
  {Pikovsky}},\ }\href@noop {} {\bibfield  {journal} {\bibinfo  {journal}
  {Phys. Rev. Lett.}\ }\textbf {\bibinfo {volume} {111}},\ \bibinfo {pages}
  {204101} (\bibinfo {year} {2013}{\natexlab{a}})}\BibitemShut {NoStop}%
\bibitem [{\citenamefont {Komarov}\ and\ \citenamefont
  {Pikovsky}(2014)}]{Komarov-Pikovsky-14}%
  \BibitemOpen
  \bibfield  {author} {\bibinfo {author} {\bibfnamefont {M.}~\bibnamefont
  {Komarov}}\ and\ \bibinfo {author} {\bibfnamefont {A.}~\bibnamefont
  {Pikovsky}},\ }\href@noop {} {\bibfield  {journal} {\bibinfo  {journal}
  {Physica D}\ }\textbf {\bibinfo {volume} {289}},\ \bibinfo {pages} {18}
  (\bibinfo {year} {2014})}\BibitemShut {NoStop}%
\bibitem [{\citenamefont {Vlasov}\ \emph {et~al.}(2015)\citenamefont {Vlasov},
  \citenamefont {Komarov},\ and\ \citenamefont
  {Pikovsky}}]{Vlasov-Komarov-Pikovsky-15}%
  \BibitemOpen
  \bibfield  {author} {\bibinfo {author} {\bibfnamefont {V.}~\bibnamefont
  {Vlasov}}, \bibinfo {author} {\bibfnamefont {M.}~\bibnamefont {Komarov}}, \
  and\ \bibinfo {author} {\bibfnamefont {A.}~\bibnamefont {Pikovsky}},\ }\href
  {http://stacks.iop.org/1751-8121/48/i=10/a=105101} {\bibfield  {journal}
  {\bibinfo  {journal} {J. Phys. A: Mathematical and Theoretical}\ }\textbf
  {\bibinfo {volume} {48}},\ \bibinfo {pages} {105101} (\bibinfo {year}
  {2015})}\BibitemShut {NoStop}%
\bibitem [{\citenamefont {L{\"u}ck}\ and\ \citenamefont
  {Pikovsky}(2011)}]{Lueck-Pikovsky-11}%
  \BibitemOpen
  \bibfield  {author} {\bibinfo {author} {\bibfnamefont {S.}~\bibnamefont
  {L{\"u}ck}}\ and\ \bibinfo {author} {\bibfnamefont {A.}~\bibnamefont
  {Pikovsky}},\ }\href@noop {} {\bibfield  {journal} {\bibinfo  {journal}
  {Physics Letters A}\ }\textbf {\bibinfo {volume} {375}},\ \bibinfo {pages}
  {2714 } (\bibinfo {year} {2011})}\BibitemShut {NoStop}%
\bibitem [{\citenamefont {Komarov}\ and\ \citenamefont
  {Pikovsky}(2013{\natexlab{b}})}]{Komarov-Pikovsky-13}%
  \BibitemOpen
  \bibfield  {author} {\bibinfo {author} {\bibfnamefont {M.}~\bibnamefont
  {Komarov}}\ and\ \bibinfo {author} {\bibfnamefont {A.}~\bibnamefont
  {Pikovsky}},\ }\href {\doibase 10.1103/PhysRevLett.110.134101} {\bibfield
  {journal} {\bibinfo  {journal} {Phys. Rev. Lett.}\ }\textbf {\bibinfo
  {volume} {110}},\ \bibinfo {pages} {134101} (\bibinfo {year}
  {2013}{\natexlab{b}})}\BibitemShut {NoStop}%
\bibitem [{\citenamefont {Komarov}\ and\ \citenamefont
  {Pikovsky}(2011)}]{Komarov-Pikovsky-11}%
  \BibitemOpen
  \bibfield  {author} {\bibinfo {author} {\bibfnamefont {M.}~\bibnamefont
  {Komarov}}\ and\ \bibinfo {author} {\bibfnamefont {A.}~\bibnamefont
  {Pikovsky}},\ }\href {\doibase 10.1103/PhysRevE.84.016210} {\bibfield
  {journal} {\bibinfo  {journal} {Phys. Rev. E}\ }\textbf {\bibinfo {volume}
  {84}},\ \bibinfo {pages} {016210} (\bibinfo {year} {2011})}\BibitemShut
  {NoStop}%
\bibitem [{\citenamefont {Buzs{\'a}ki}(2006)}]{Buzsaki-06}%
  \BibitemOpen
  \bibfield  {author} {\bibinfo {author} {\bibfnamefont {G.}~\bibnamefont
  {Buzs{\'a}ki}},\ }\href@noop {} {\emph {\bibinfo {title} {Rhythms of the
  brain}}}\ (\bibinfo  {publisher} {Oxford UP},\ \bibinfo {address} {Oxford},\
  \bibinfo {year} {2006})\BibitemShut {NoStop}%
\bibitem [{\citenamefont {Rosjat}\ \emph {et~al.}(2014)\citenamefont {Rosjat},
  \citenamefont {Popovych},\ and\ \citenamefont
  {Daun-Gruhn}}]{Rosjat-Popovych-Daun-14}%
  \BibitemOpen
  \bibfield  {author} {\bibinfo {author} {\bibfnamefont {N.}~\bibnamefont
  {Rosjat}}, \bibinfo {author} {\bibfnamefont {S.}~\bibnamefont {Popovych}}, \
  and\ \bibinfo {author} {\bibfnamefont {S.}~\bibnamefont {Daun-Gruhn}},\
  }\href@noop {} {\bibfield  {journal} {\bibinfo  {journal} {Theoretical
  Biology and Medical Modelling}\ }\textbf {\bibinfo {volume} {11}} (\bibinfo
  {year} {2014})}\BibitemShut {NoStop}%
\bibitem [{\citenamefont {Komarov}\ \emph {et~al.}(2014)\citenamefont
  {Komarov}, \citenamefont {Gupta},\ and\ \citenamefont
  {Pikovsky}}]{Komarov-Gupta-Pikovsky-14}%
  \BibitemOpen
  \bibfield  {author} {\bibinfo {author} {\bibfnamefont {M.}~\bibnamefont
  {Komarov}}, \bibinfo {author} {\bibfnamefont {S.}~\bibnamefont {Gupta}}, \
  and\ \bibinfo {author} {\bibfnamefont {A.}~\bibnamefont {Pikovsky}},\ }\href
  {http://stacks.iop.org/0295-5075/106/i=4/a=40003} {\bibfield  {journal}
  {\bibinfo  {journal} {EPL}\ }\textbf {\bibinfo {volume} {106}},\ \bibinfo
  {pages} {40003} (\bibinfo {year} {2014})}\BibitemShut {NoStop}%
\end{thebibliography}
\end{document}